\documentclass[final,5p,times,twocolumn]{elsarticle}

\usepackage{amssymb,amsmath}
\usepackage{algorithm}
\usepackage{algorithmic}
\usepackage{multirow}
\usepackage{appendix}
\usepackage{url}
\usepackage{caption} 
\usepackage{setspace}
\usepackage{algorithm}
\usepackage{algorithmic}
\usepackage{natbib}
\usepackage{makecell}

\usepackage{lineno}
\usepackage{float}

\journal{Computerized Medical Imaging and Graphics}

\makeatletter
\def\ps@pprintTitle{%
 \let\@oddhead\@empty
 \let\@evenhead\@empty
 \def\@oddfoot{}%
 \let\@evenfoot\@oddfoot}
\makeatother

\begin{document}

\begin{frontmatter}

\title{Whole brain segmentation with full volume neural network}

\author[1]{Yeshu Li\fnref{fn1}}
\ead{yli299@uic.edu}
\author[6]{Jonathan Cui\fnref{fn1}}
\ead{jonathancui03@gmail.com}
\author[3,4]{Yilun Sheng\fnref{fn1}}
\ead{ridic2651@gmail.com}
\author[2]{Xiao Liang}
\ead{liangx@rdfz.cn}
\author[4]{Jingdong Wang}
\ead{jingdw@microsoft.com}
\author[4]{Eric I-Chao Chang}
\ead{echang@microsoft.com}
\author[4,5]{Yan Xu\corref{cor1}}
\ead{xuyan04@gmail.com}
\fntext[fn1]{This work was done when Yeshu, Jonathan and Yilun were interns at Microsoft Research Asia in 2019.}
\cortext[cor1]{Corresponding author}

\fntext[fnt]{$\copyright$ 2021. This manuscript version is made available under the CC-BY-NC-ND 4.0 license \url{https://creativecommons.org/licenses/by-nc-nd/4.0/}. This manuscript is accepted to Computerized Medical Imaging and Graphics. DOI: \url{https://doi.org/10.1016/j.compmedimag.2021.101991}.}

\address[5]{School of Biological Science and Medical Engineering and Beijing Advanced Innovation Centre for Biomedical Engineering, Beihang University, Beijing 100191, China}
\address[1]{Department of Computer Science, University of Illinois at Chicago, Chicago, IL 60607, United States}
\address[6]{Vacaville Christian Schools, Vacaville, CA 95687, United States}
\address[2]{High School Affiliated to Renmin University Of China, Beijing 100080, China}
\address[3]{Institute for Interdisciplinary Information Sciences, Tsinghua University, Beijing 100084, China}
\address[4]{Microsoft Research, Beijing 100080, China}

\begin{abstract}
Whole brain segmentation is an important neuroimaging task that segments the whole brain volume into anatomically labeled regions-of-interest. Convolutional neural networks have demonstrated good performance in this task. Existing solutions, usually segment the brain image by classifying the voxels, or labeling the slices or the sub-volumes separately. Their representation learning is based on parts of the whole volume whereas their labeling result is produced by aggregation of partial segmentation. Learning and inference with incomplete information could lead to sub-optimal final segmentation result. To address these issues, we propose to adopt a full volume framework, which feeds the full volume brain image into the segmentation network and directly outputs the segmentation result for the whole brain volume. The framework makes use of complete information in each volume and can be implemented easily. An effective instance in this framework is given subsequently. We adopt the $3$D high-resolution network (HRNet) for learning spatially fine-grained representations and the mixed precision training scheme for memory-efficient training. Extensive experiment results on a publicly available $3$D MRI brain dataset show that our proposed model advances the state-of-the-art methods in terms of segmentation performance.
\end{abstract}


\end{frontmatter}


\section{Introduction}
\label{sec:introduction}

Magnetic resonance imaging (MRI) of the brain has been widely adopted in various neuroimaging studies and medical practice since it provides a fast, noninvasive, safe and painless way of acquiring good-contrast anatomical information of the human brain. Segmentation of neuroanatomy is a fundamental task in MRI analysis where voxel-wise classification is performed on a brain image by assigning each voxel a semantic label representing a specific neuroanatomical structure such that voxels belonging to the same category induce similar features. As a basic tool of granting access to quantitative measurements like volume, thickness and shape of neuroanatomy from MRI, brain segmentation is of significance to clinical applications including brain development monitoring, brain morphological analysis, deep brain stimulation and pre-operative evaluation \citep{gonzalez2016review}. Usually, brain structure segmentation focuses on one or several brain structures. In contrast, whole brain segmentation aims for more than tens of regions, thus more challenging. Detailed whole brain segmentation is particularly useful in fine-grained quantitative brain study such as structural brain network analysis and brain connectivity analysis \citep{huo20193d}. Manual segmentation of brain MRI following some human brain labeling protocol serves as a golden standard in practice, but is laborious to obtain from human raters with expert level medical knowledge, and poorly reproducible because of operator variability. As a result, accurate fully automatic whole brain segmentation methods are highly desirable and have attracted emerging interest of research communities for a long period of time.

Atlas-based approaches are conventional automatic segmentation methods based on deformable registration and a manually segmented atlas. FreeSurfer \citep{fischl2012freesurfer} and BrainSuite \citep{shattuck2002brainsuite} are well-known software tools for atlas-based whole brain segmentation, which have inspired multi-atlas label fusion approaches \citep{wang2013multi, asman2013non} that have demonstrated state-of-the-art segmentation accuracy. The segmentation performance of these methods is highly dependent on the quality of reference atlases, alignment estimation and fusion strategies. Another class of automatic whole brain segmentation methods is based on machine learning classifiers \citep{zhang2001segmentation, tu2009auto}. These approaches rely on a predefined set of hand-crafted feature representations which are usually sub-optimal to target applications and inefficient to be extracted.

Deep learning methods with convolutional neural networks (CNNs) \citep{krizhevsky2012imagenet} and fully convolutional networks (FCNs) \citep{long2015fully} have shown state-of-the-art performance in computer vision tasks. There have been an increasing number of CNNs/FCNs proposed for whole brain segmentation in the past few years, as a result of their ability to learn task-specific relevant features and classifiers at the same time. Most existing works focus on automatic brain segmentation methods that are patch-based \citep{fang2019automatic}, slice-based \citep{chen2018drinet, roy2019quicknat} or sub-volume-based \citep{huo20193d, sun20193d}. A remarkable drawback of these approaches is that only partial information of each data instance is adopted during training and inference. Henceforth, they are susceptible to underfitting the data in view of the fact that only a subset of the available intensity information is provided for training. Moreover, a hand-crafted fusion strategy for aggregating partial prediction results is necessary. The above approaches are based on partial volumes mainly because of the difficulty in training 3D neural networks and the characteristics of volumetric brain MRI datasets. More specifically, on typical graphics processing unit (GPU) platforms, if trained with complete whole brain volumes without cropping, a CNN/FCN would become too shallow to be useful and usually exhibit performance inferior to multi-atlas methods. Furthermore, most of the proposed CNNs/FCNs were validated on a brain segmentation task with only a few brain structures. Only a few deep learning-based approaches evaluated their whole brain segmentation performance on datasets with more than $50$ anatomical labels \citep{henschel2020fastsurfer, coupe2020assemblynet, li2017compactness, fang2019automatic, huo20193d}.

A full volume training scheme that makes use of complete whole brain data as input possesses attractive properties. First, learning and inference with original volumes result in no loss of available information in the input data. Second, a full volume approach frees ourselves from designing sophisticated divide-and-aggregate schemes for volumes. Third, the inference time (excluding preprocessing) for the input volume becomes approximately milliseconds because of only one typical neural network forward pass. Fourth, an FCN admits input volumes with no hard shape constraints, able to integrate advanced FCN techniques more seamlessly.

In spite of the above advantages, a full volume training framework gives rise to three major challenges. First of all, large-volume brain MRI data makes common 3D FCN architectures prohibitively expensive to be deployed in current GPUs. Second, the FCN architecture itself is required to be effective in segmenting fine-grained brain structures, and compact for limited GPU resources. The last challenge is concerned with the scarcity of free manually labeled brain MRI scans. Note that partial volume-based approaches have access to plenty of training data due to their sampling procedure while a full volume framework does not.

In this paper, we propose to adopt a full volume framework for whole brain segmentation in practical use and instantiate it with effective components to resolve the above issues. This is one of the early attempts to use full brain volumes with a deep learning model for whole brain segmentation. A high-resolution FCN with accurate brain structure delineation ability and low GPU memory usage is introduced, consisting of three sub-networks. The stem sub-network at the beginning and the regression sub-network in the end are leveraged for feature map size reduction and voxel-wise classification respectively. The backbone sub-network in the middle includes several stages retaining different abstract levels of feature maps. Residual connections are incorporated to strengthen gradient flow, whereas dense fusion connections are adopted to facilitate feature reusability. Along with that, we alleviate the GPU resource intensiveness issue by taking advantage of the advances in reduced-precision deep neural network (DNN) training, which not only reduces the maximum allocated GPU memory but also yields potential computation acceleration. Extensive experiments conducted on a publicly available brain image dataset demonstrate that our proposed method achieves significant improvement over state-of-the-art approaches in terms of overall segmentation accuracy and is about $200$ times faster than traditional methods regarding inference time excluding preprocessing. We further validate the effectiveness of mixed precision training, high-resolution FCN and volumetric image augmentation by conducting a series of ablation studies. The source code and models are publicly available\footnote{\url{https://github.com/microsoft/VoxHRNet}}.

The main contributions of this work can be summarized as follows:

\begin{itemize}
    \item We propose a full volume method, different from patch-based and sub-volume-based methods. These partial-volume-based methods are slow and hindered by incomplete information. The benefit of our method is from training and segmenting a $3$D volume directly. The framework is simple and general, in which any structured prediction models that produce correspondingly-sized output of the input can be incorporated.
    \item We adopt a high-resolution FCN in our framework for whole brain segmentation. Compared to traditional FCNs, the adopted FCN learns spatially rich representation by retaining all levels of features (original and downsampled) and facilitating feature reuse via dense fusion throughout the network.
    \item We enable our full volume network to be trained on 12GB GPUs by mixed precision training, which alleviates the critical issue regarding large GPU memory. The resulting inference time (excluding preprocessing) is about $200$ times less than traditional methods (FCN \citep{long2015fully}, U-Net \citep{ronneberger2015u}, FastSurfer \citep{henschel2020fastsurfer}). It outperforms the state-of-the-art (MA-FCN \citep{fang2019automatic}) by about $2\%$ absolute improvement in Dice overlap.
\end{itemize}

\section{Related work}
\label{sec:relatedwork}


\subsection{Brain tissue segmentation}

The task of segmenting brain tissue mainly refers to partitioning a brain image into four regions including white matter (WM), gray matter (GM), cerebrospinal fluid (CSF) and non-brain tissue \citep{dora2017state, valverde2015comparison, liew2006current}. Brain tissue segmentation provides a quantitative tool for brain morphological analysis, surgical planning, disease detection and a wide range of clinical applications. Region-based approaches \citep{xue2001knowledge, gui2012morphology} take advantage of the homogeneity property of similar voxels in the same region for boundary detection. Contour-based deformable models \citep{huang2009hybrid, kapur1996segmentation}, region growing methods \citep{tang2000mri}, level set methods \citep{wang2010level, chen2008fuzzy, wang2013longitudinally} and graph-based approaches \citep{song2006integrated} all fall under the region-based category. Thresholding technique \citep{kalavathi2013brain, de2009white, sathya2011optimal, kapur1985new} is another class of segmentation methods that labels a voxel by comparing its intensity value with a local or global threshold, which is nonetheless sensitive to statistical fluctuation such as noise, artifacts, multi-modality and intensity heterogeneity. Unsupervised learning methods \citep{constantin2010unsupervised} such as clustering were proposed for brain tissue segmentation as well, in which fuzzy c-means (FCM) \citep{barra2000tissue, shen2005mri, halder2019brain} that progressively refines a soft-clustering membership function for each voxel, and mixture models \citep{rajapakse1996technique, da2007dirichlet, kouw2019cross, yousefi2012brain, zhang2001segmentation, wells1996adaptive, van1999automated, ashburner2005unified, liang1994parameter} that estimate the clusters with a statistical model by maximum likelihood (ML) or maximum a posteriori (MAP), are some of the popular clustering approaches. Moreover, training a supervised machine learning classifier with a set of hand-crafted features extracted from the images is another important line of work \citep{kong2014discriminative, paul2015automated, vrooman2007multi, anbeek2005probabilistic, anbeek2008probabilistic, van2013automated, yi2009discriminative}. In addition, brain tumor segmentation \citep{liu2014survey} has been widely studied for differentiating abnormal tissues such as edema and necrotic core from normal brain tissues.

\subsection{Brain structure segmentation}

Brain structure segmentation methods, usually developed to segment a group of structures, such as subcortical structures, or a specific neuroanatomical structure, possibly with its substructures, are challenging and non-trivial on account of the obscure boundaries among different classes of structures given only MR image intensity information \citep{gonzalez2016review}. Morphometric analysis is one of the important clinical applications of brain structure segmentation because brain morphological characteristics change is shown to be closely related to neuropsychiatric and neurodegenerative disorders \citep{chudasama2006functions}. For instance, hippocampus atrophy has been demonstrated to be a key feature of Alzheimer's disease and epilepsy \citep{apostolova2010subregional}, while quantitative evaluation of the area of the substantia nigra in the brainstem has been proposed as a possible technique for diagnosing Parkinson's disease \citep{sakalauskas2010transcranial}. There is a rich body of work on segmentation methods for a few brain structures \citep{kushibar2018automated}. A knowledge-driven algorithm combining the shape information in an initially identified lateral ventricle and some anatomical knowledge was proposed by \citet{xia2007automatic} to segment the caudate nucleus region in a human brain MR image. For the whole hippocampus segmentation as well as its subfields, \citet{pipitone2014multi} came up with a general automatic template-generating framework that efficiently makes use of a small atlas library.

\subsection{Traditional whole brain segmentation}

Rather than extracting only a few regions, whole brain segmentation \citep{fischl2002whole} typically performs a brain segmentation task by assigning each voxel a label from a predefined set containing at least tens of neuroanatomical structures. Delineation of a large number of brain structures all at once is non-trivial for volumetric analysis and sometimes provides better segmentation performance than individual brain structure segmentation, while it is challenging because of the large number of labels as well as similar intensity histograms of different structures \citep{fischl2002whole}. FreeSurfer \citep{fischl2012freesurfer}, whose segmentation method is learning-based and atlas-based, and FIRST \citep{patenaude2011bayesian}, based on a deformable appearance model, are widely adopted whole brain segmentation softwares. Joint label fusion (JLF) with corrective learning, developed by \citet{wang2013multi}, is the one of the well-known multi-atlas whole brain segmentation methods that won the first place in the 2012 MICCAI Multi-Atlas Labeling Challenge \citep{landman2012miccai}. In the mean time, \citet{asman2013non}, integrated a rater performance level estimation process and a non-local correspondence perspective into a statistical label fusion framework, showing better performance than baseline algorithms including locally weighted voting methods \citep{sabuncu2010generative} and Spatial STAPLE \citep{commowick2012estimating}. Thereafter, \citet{huo2016consistent} came up with Multi-atlas CRUISE that achieves self-consistent whole brain segmentation and cortical surface reconstruction results, resolving the spatially inconsistent issue caused by independent modeling of the two tasks. To make an automated brain segmentation method robust to various imaging protocols across different acquisition platforms, \citet{puonti2016fast} presented an unsupervised Gaussian mixture model along with a mesh-based probabilistic atlas to produce sequence-independent segmentation results.

\subsection{Deep learning for whole brain segmentation}

There has been an increasing interest in developing CNNs/FCNs for whole brain segmentation recently. \citet{fang2019automatic} proposed a patch-based multi-atlas guided FCN (MA-FCN) consisting of three pathways of target and atlas patches. A slice-based framework named QuickNAT was put forward by \citet{roy2019quicknat}, who suggested a pretrain-and-finetune framework taking advantage of a large amount of unlabeled brain volumes and off-the-shelf automatic segmentation softwares. Concurrently, \citet{huo20193d} developed a spatially localized atlas network tiles (SLANT) method, which is sub-volume-based, where multiple spatially location-specific FCNs are trained in parallel, with auxiliary labels generated by Non-Local Spatial STAPLE (NLSS) \citep{asman2014hierarchical}. Other sub-volume-based automatic brain segmentation methods in the literature include HighRes3DNet \citep{li2017compactness}, VoxResNet \citep{chen2018voxresnet}, SW-3D-UNet \citep{sun20193d} and HyperDense-Net \citep{dolz2018hyperdense}. DRINet \citep{chen2018drinet} falls under the slice-based category, whereas DeepNAT \citep{wachinger2018deepnat} belongs to 3D patch-based methods.

The vast majority of existing works on automatic whole brain segmentation with FCNs are either slice-based or sub-volume-based due to the increasing memory demand for training. In other words, there are few works training their models with complete volumes in an end-to-end manner. There is similar work concurrent to ours, all of which are the early attempts to adopt complete volumes in both training and testing. A 3D U-Net-like FCN trained with full volumes was adopted by \citet{billot2020a} to perform a contrast-agnostic segmentation task involving $37$ regions-of-interest (ROIs), where the authors focus on a novel augmentation method based on a generative model. By contrast, our work emphasizes on practically solving the whole brain segmentation task for much more brain structures with a full volume framework and adopting a more advanced FCN architecture, the high-resolution segmentation network, to deal with accurate segmentation of small neuroanatomical regions.

\subsection{Deep learning network architectures for semantic segmentation}

CNNs \citep{krizhevsky2012imagenet} are deep learning techniques that have achieved great success in computer vision applications including semantic segmentation \citep{girshick2014rich}. Without fully connected layers, FCNs \citep{long2015fully} were proposed to realize end-to-end pixel-wise dense prediction for input images with arbitrary sizes. From the perspective of representation learning for semantic segmentation, one line of works \citep{chen2017deeplab, chen2016attention, zhao2017pyramid} extended FCNs to produce coarse segmentation maps based on low-resolution or medium-resolution representations. Another family of architectures recover high-resolution representations by upsampling low-resolution ones. The upsample sub-network can be symmetric to the downsample FCN, such as U-Net \citep{ronneberger2015u}, V-Net \citep{milletari2016v}, SegNet \citep{badrinarayanan2017segnet} and DeconvNet \citep{noh2015learning}, or asymmetric, for example, RefineNet \citep{lin2017refinenet}, stacked DeConvNet \citep{fu2019stacked} and DeepLabv3+ \citep{chen2018encoder}. Maintaining high-resolution representations throughout the forward passing process is key to learning strong segmentation maps \citep{fourure2017residual, saxena2016convolutional}. Multi-scale fusion facilitates information exchange across representations in different resolutions via skip connection or pyramid pooling \citep{badrinarayanan2017segnet, zhao2017pyramid, xie2018interleaved, zhou2019unetplusplus, zhou2018unetplusplus}. \citet{wang2020deep} proposed a general network architecture paradigm for deep high-resolution representation learning that achieved great success in various visual recognition tasks by maintaining high-resolution representations while imposing dense multi-scale fusion.

\subsection{Reduced-precision training for neural networks}

There have been works on training CNNs with binary activations \citep{hubara2016binarized}, and training recurrent neural networks (RNNs) with quantized gradients \citep{hubara2017quantized}. \citet{micikevicius2018mixed} proposed a general-purpose mixed precision training framework for training large-scale DNNs efficiently, almost halving the GPU memory usage. A mixed precision training framework adopting BFLOAT16 format, able to represent the same range of values as FP32 is, was presented by \citet{kalamkar2019study} to avoid the entailment of loss scaling in \citep{micikevicius2018mixed}. Recently, \citet{yang2019swalp} proposed a low-precision stochastic gradient descent (SGD) approach by taking advantage of stochastic weight averaging and quantizing the gradient accumulator as well as the velocity vector. In order to determine the quantization precision of each layer in a systematic way and avoid the large search space, \citet{dong2019hawq} suggested using the second-order information in each block of the network. In addition, a fully half-precision training strategy put forward by \citet{cheng2019distributed} was shown to possess no sacrificed accuracy compared to single precision in large-scale distributed training. Above all, reduced-precision training is expected to benefit a deep learning model for large-volume semantic segmentation.

\subsection{Image augmentation}

Data augmentation comes from a priori invariance property of the data distribution and serves as a regularization technique to combat model overfitting. \citet{simard2003best} illustrated a successful practice of image data augmentation by postulating elastic distortion invariance in a handwritten digit dataset. Various image augmentation strategies were adopted in AlexNet \citep{krizhevsky2012imagenet} on the large-scale ImageNet dataset \citep{deng2009imagenet}. The effectiveness of data augmentation for image classification was studied by \citet{perez2017effectiveness}. \citet{hauberg2016dreaming} proposed to learn class-dependent augmentation schemes in which transformation is defined in a finite-dimensional Lie group of diffeomorphisms. Furthermore, image augmentation has been applied in medical image analysis based on radial transform \citep{salehinejad2018image}, appearance models \citep{zhao2019data} and generative adversarial networks (GANs) \citep{pandey2020image}. Augmentation plays an important role in handling scarce data such as typical brain MRI datasets.

\section{Full volume brain segmentation framework}
\label{sec:fullvolume}

We propose to adopt a full volume framework for brain segmentation in this work. Compared to most existing deep learning methods, the framework makes prediction for each full volume in a holistic, faster and more straightforward way. We start by discussing key properties of a successful brain segmentation neural network approach and introducing existing paradigms, after which we present the formulation of the framework.

Some important properties can be identified for a system based on neural networks to be successful in the challenging brain segmentation task: (1) the system should be easy to understand and implement, which makes it reproducible and generalizable; (2) the network model is supposed to be fed with as much training data as possible because of the nature of neural networks and training via backpropagation; (3) a practical amount of compute resource is allowed so that the system is deployable in common platforms; (4) a rich set of hierarchical features capturing local and global contexts is necessary for accurate voxel-wise prediction; (5) the system must be fast enough to be applicable in clinical use, for example, aiding pathologists in morphological analysis or neuroanatomical structure localization. Most existing neural network-based approaches do not possess all the above properties. The full volume framework fulfills all the requirements, mainly because it directly tackles whole brain volumes and deals with subsequent challenges accordingly.

\subsection{Brain segmentation neural networks with partial volumes}

In a gray-scale volumetric image semantic segmentation task with $L$ target classes, given a training set of $n$ samples, where $X_i \in \mathbb{R}^{H_i \times W_i \times D_i}$, $Y_i \in \{1, \dots, L\}^{H_i \times W_i \times D_i}$ denote the intensity map and the label map for the $i$-th training volume respectively with $H_i$, $W_i$ and $D_i$ standing for its dimensions, the goal is to learn a function, or a neural network model, $F_\theta$ parameterized by $\theta$ that maps a target image $X_t \in \mathbb{R}^{H_t \times W_t \times D_t}$ to its label map prediction $\hat{Y}_t \in \{1, \dots, L\}^{H_t \times W_t \times D_t}$.

Gradient backpropagation with GPU enables efficient learning. However, a common brain volume usually encompasses over millions of voxels without down-sampling, which makes it challenging to train a completely end-to-end neural network and maintain sufficient model capacity with GPU-enabled backpropagation.

On account of the above limitations, most of the state-of-the-art whole brain segmentation methods train their deep learning models on a set of cropped data from original whole brain volumes. The inference for a target volume is thus performed by aggregating the segmentation results of sampled sub-volumes into a complete label map. Patch-based segmentation methods classify each voxel or its 3D neighborhood \citep{dolz20183d} based on an extracted patch enclosing them, commonly through a CNN with fully connected layers in the end or an FCN respectively. The prediction and aggregation procedure can be formulated as

\begin{equation}
    \hat{Y} = \Phi(\{F_\theta(F_p(X, v)) | v \in \Omega\}),
\end{equation}
where $\Phi(\cdot)$ is an aggregator function, $F_p(\cdot, \cdot)$ is a patch extraction function and $\Omega = \{1, \dots, HWD\}$ denotes the set of indices for each voxel in volume $X$. On the contrary, slice-based methods adopt FCNs to obtain the segmentation result of each input slice, and synthesize into a label volume the outputs for all the 2D slices cut off along different axes. More formally,

\begin{equation}
    \hat{Y} = \Phi(\{F_\theta(X_j^d) | d \in \{H, W, D\}, 1 \leq j \leq d\}),
\end{equation}
where $X_j^H \in \mathbb{R}^{W \times D}$, $X_j^W \in \mathbb{R}^{H \times D}$ and $X_j^D \in \mathbb{R}^{H \times W}$ represent the $j$-th slice along the first, second and third dimension of $X$ respectively. Similarly, prediction of models trained on sub-volumes can be written as

\begin{equation}
    \hat{Y} = \Phi(\{F_\theta(X') | X' \in S, S \subseteq \mathcal{P}(X)\}),
\end{equation}
where $\mathcal{P}(X)$ denotes the set of all the sub-volumes of $X$ and $S$ can be carefully designed to serve as a sub-volume dividing strategy or randomly sampled from $\mathcal{P}(X)$.

\begin{figure}[!t]
    \centering
    \includegraphics[width=0.7\columnwidth]{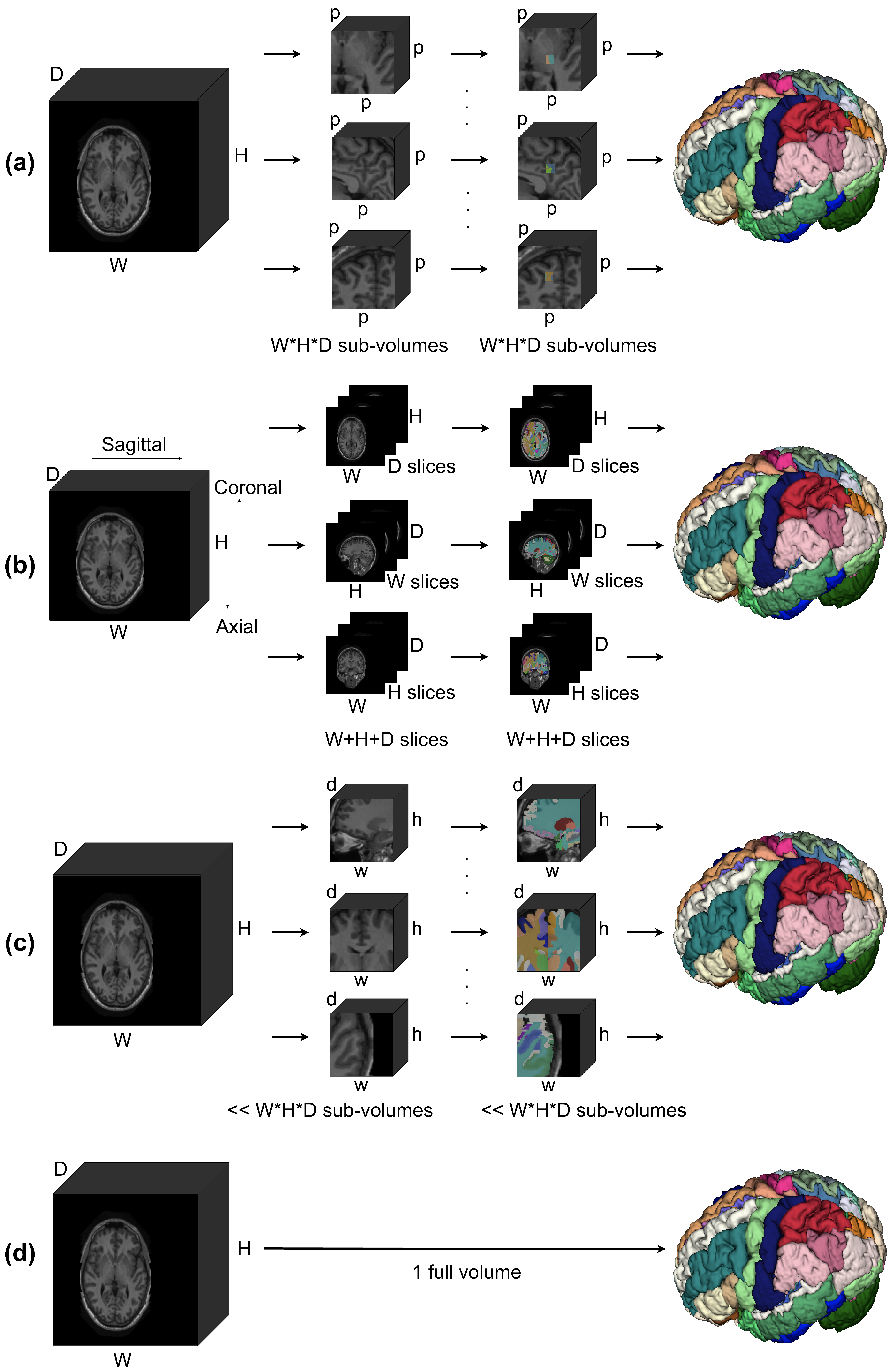}
    \caption{Comparisons of different training approaches. $W$, $H$ and $D$ denote volume dimensions. Ellipses indicate that there are rows omitted for illustration purpose. (a) patch-based (predicting the labels of patch center) method. $p$ is the patch size. (b) 2D slice-based (predicting the labels of 2D slices) method. (c) sub-volume-based (predicting the labels of 3D sub-volumes) method. $w$, $h$ and $d$ are dimensions of sampled sub-volumes. (d) full-volume-based (predicting the labels of complete 3D volumes). Best viewed in color.}
    \label{fig:end2endcomp}
\end{figure}

A noteworthy drawback of the above approaches is that many intermediate partial segmentation results have to be produced during the process based on local regions, which is usually time-consuming in comparison with the time of only one forward pass. Another shortcoming is that these approaches entail a strategy of acquiring and aggregating partial prediction results from the sampled local regions.

\subsection{Formulation}
\label{sec:formulation}

We avoid the limitations of existing methods by proposing to train a fully convolutional network holistically on whole brain volumes. Formally, the full volume model yields a predicted output $\hat{Y}$ for an input volume $X$ as follows:

\begin{equation}
    \hat{Y} = F_\theta (X).
\end{equation}
An illustration of the full volume framework and the three divide-and-aggregate methods is given in Figure \ref{fig:end2endcomp}. By full volume, we emphasize on the whole brain volumes adopted in both training and inference. Any instance in the full volume framework leads to a completely image-to-image segmentation system, while the patch-to-pixel or patch-to-patch strategies adopted by most existing methods normally result in inferior learning and inference ability. This also significantly reduces the framework complexity because there is no need to develop fusion strategy for partial segmentation results. As shown in Figure \ref{fig:end2endcomp}, it can be observed that our framework requires only one forward pass and no sampling strategy, which implies that it naturally spends less than one second on forward passing for one brain volume with GPUs.

Now we can make use of complete ground truth label maps $Y_i$ directly to train a model in the full volume framework via empirical risk minimization (ERM) and backpropagation:

\begin{equation}
    \theta^* = \arg \min_\theta \frac{1}{n} \sum_{i = 1}^n \mathcal{L}(F_\theta(X_i), Y_i),
\end{equation}
where $\mathcal{L}(\cdot, \cdot)$ is a loss function that measures the discrepancy between prediction and the desired output.

We consider three loss functions as training objectives:

\begin{equation}
    \begin{split}
    & \mathcal{L}_{CE} = - \frac{1}{|B|} \sum_{i \in B} \frac{1}{|\Omega_i|} \sum_{v \in \Omega_i} \sum_{l = 1}^L Y_i^l(v) \log(\hat{P}_i^l(v)), \\
    & \mathcal{L}_{Dice} = 1 - \frac{1}{L} \frac{1}{|B|} \sum_{l = 1}^L \sum_{i \in B}\frac{\epsilon + 2 \sum_{v \in \Omega_i} Y_i^l(v) \hat{P}_i^l(v)}{\epsilon + \sum_{v \in \Omega_i}(Y_i^l(v) + \hat{P}_i^l(v))}, \\
    & \mathcal{L}_{Comb} = \mathcal{L}_{CE} + \mathcal{L}_{Dice}, \\
    \end{split}
\end{equation}
where $B \subseteq \{1, \dots, n\}$ represents a mini-batch and $\Omega_i \subset \mathbb{Z}$ stands for the voxel indices set of the $i$-th image with $\epsilon$ being an additive smoothness constant. $Y_i^l(v)$ denotes the ground truth probability of voxel $v$ being labeled as $l$ in the $i$-th image, and $\hat{P}_i^l(v)$ is the corresponding estimated probability. For crisp segmentation in our case, every voxel is mapped to exactly one label in the ground truth with probability $1$. $\mathcal{L}_{CE}$ is the standard multi-class cross entropy loss, $\mathcal{L}_{Dice}$ is the dice loss and $\mathcal{L}_{Comb}$ is a combined loss as an unweighted sum of the above two.


\section{A full volume segmentation network instance}

\subsection{Network architecture}

In order to learn spatially rich representation and produce accurate segmentation especially for small neuroanatomical regions, inspired by \citet{sun2019deep, sun2019high, wang2020deep}, we introduce a high-resolution network named VoxHRNet that fits in the full volume framework.

\begin{figure*}[!t]
    \centering
    \includegraphics[width=\textwidth]{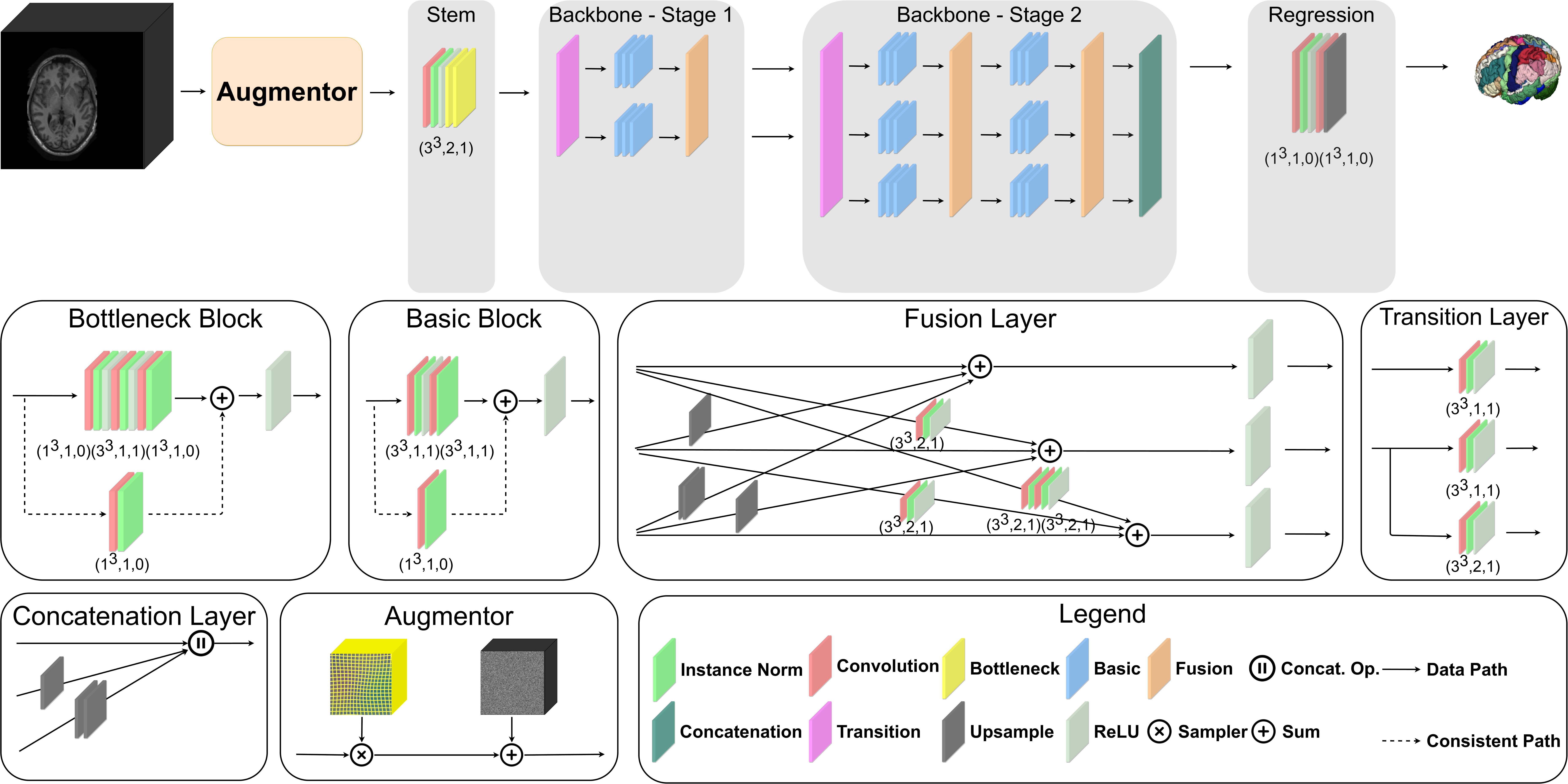}
    \caption{The network architecture of our proposed high-resolution FCN model. The consistent path goes through a convolution layer together with an instance normalization layer whenever the number of input channels does not match the desired number of output channels to resolve inconsistency. It will be replaced by a normal data path when a residual connection can be placed without channel inconsistency. The parameters $(a^3, b, c)$ below each convolution layer indicate that it has a kernel size of $a \times a \times a$, a stride of $b$ and a padding size of $c$. Best viewed in color.}
    \label{fig:architecture}
\end{figure*}

\begin{table*}[!t]
    \renewcommand{\arraystretch}{1.3}
    \caption{The detailed network architecture of our proposed high-resolution FCN model. The first two components are important building blocks of the networks, whereas the architecture description is from the fourth row to the last row sequentially as indicated by the texts in brackets. The sequential concatenation operation is denoted by $\parallel$, the sum operation is denoted by $+$ and the repetition operation is denoted by $*$. The number of intermediate channels is set accordingly and can be inferred from the context. The number at the beginning of each row for branches implies the next branch number to which the data flows. The dimension sizes are rounded to integral values in practice. Interpolation is applied if upsampling results in dimension inconsistency. Conv($a, b, c$) indicates that the convolutional layer has a kernel size of $a \times a \times a$, a stride of $b$ and a padding size of $c$.}
    
    \centering
    \resizebox{\textwidth}{!}{
        \begin{tabular}{cccccc}
            \hline
            Component & Output dimension & Output channels & Branch 1 & Branch 2 & Branch 3 \\
            \hline
            Bottleneck & Same as input & Intermediate * 4 & \makecell{1: \{[Conv(1, 1, 0) $\parallel$ Norm $\parallel$ ReLU $\parallel$ Conv(3, 1, 1) $\parallel$ Norm $\parallel$ ReLU \\ $\parallel$ Conv(1, 1, 0) $\parallel$ Norm] + [Conv(1, 1, 0) $\parallel$ Norm]\}  $\parallel$ ReLU} & & \\
            \hline
            Basic & Same as input & Same as intermediate & \makecell{1: \{[Conv(3, 1, 1) $\parallel$ Norm $\parallel$ ReLU $\parallel$ Conv(3, 1, 1) $\parallel$ Norm] \\ + [Conv(1, 1, 0) $\parallel$ Norm]\}  $\parallel$ ReLU} & & \\
            \hline
            Input (begin) &	W * H * D &	1 & & & \\
            \hline
            Stem &	W/2 * H/2 * D/2 & 64 & \makecell{1: Conv(3, 2, 1) $\parallel$ Norm $\parallel$ ReLU $\parallel$ Bottleneck * 2} & & \\
            \hline
            Transition 1 & \makecell{W/2 * H/2 * D/2 \\ W/4 * H/4 * D/4} & \makecell{16 \\ 32} & \makecell{1: Conv(3, 1, 1) $\parallel$ Norm $\parallel$ ReLU \\ 2: Conv(3, 2, 1) $\parallel$ Norm $\parallel$ ReLU} & & \\
            \hline
            Stage 1 & \makecell{W/2 * H/2 * D/2 \\ W/4 * H/4 * D/4} & \makecell{16 \\ 32} & \makecell{1: Basic * 3} & \makecell{2: Basic * 3} & \\
            \hline
            Fusion 1 & \makecell{W/2 * H/2 * D/2 \\ W/4 * H/4 * D/4} & \makecell{16 \\ 32} & \makecell{1: Identity \\ 2: Conv(3, 2, 1) $\parallel$ Norm $\parallel$ ReLU} & \makecell{1: Upsample \\ 2: Identity} & \\
            \hline
            Transition 2 & \makecell{W/2 * H/2 * D/2 \\ W/4 * H/4 * D/4 \\ W/8 * H/8 * D/8} & \makecell{16 \\ 32 \\ 64} & \makecell{1: Conv(3, 1, 1) $\parallel$ Norm $\parallel$ ReLU} & \makecell{2: Conv(3, 1, 1) $\parallel$ Norm $\parallel$ ReLU \\ 3: Conv(3, 2, 1) $\parallel$ Norm $\parallel$ ReLU} & \\
            \hline
            Stage 2.1 & \makecell{W/2 * H/2 * D/2 \\ W/4 * H/4 * D/4 \\ W/8 * H/8 * D/8} & \makecell{16 \\ 32 \\ 64} & \makecell{1: Basic * 3} & \makecell{2: Basic * 3} & \makecell{3: Basic * 3} \\
            \hline
            Fusion 2.1 & \makecell{W/2 * H/2 * D/2 \\ W/4 * H/4 * D/4 \\ W/8 * H/8 * D/8} & \makecell{16 \\ 32 \\ 64} & \makecell{1: Identity \\ 2: Conv(3, 2, 1) $\parallel$ Norm $\parallel$ ReLU \\ 3: [Conv(3, 2, 1) $\parallel$ Norm] * 2 $\parallel$ ReLU} & \makecell{1: Upsample \\ 2: Identity \\ 3: Conv(3, 2, 1) $\parallel$ Norm $\parallel$ ReLU} & \makecell{1: Upsample * 2 \\ 2: Upsample \\ 3: Identity} \\
            \hline
            Stage 2.2 & \multicolumn{5}{c}{Same as Stage 2.1} \\
            \hline
            Fusion 2.2 & \multicolumn{5}{c}{Same as Fusion 2.1} \\
            \hline
            Concatenation & W/2 * H/2 * D/2 & 112 & \makecell{1: Identity} & \makecell{1: Upsample} & \makecell{1: Upsample * 2} \\
            \hline
            Regression (end) & W * H * D & 55 & \makecell{1: Conv(1, 1, 0) $\parallel$ Norm $\parallel$ ReLU $\parallel$ Conv(1, 1, 0) $\parallel$ Upsample} & & \\
            \hline
        \end{tabular}
    }
    \label{tab:architecture}
\end{table*}

The architecture of VoxHRNet is illustrated in Figure \ref{fig:architecture} with a layer-wise description in Table \ref{tab:architecture}. Note that the architecture illustrated at the top of Figure \ref{fig:architecture} from left to right corresponds to the sequence of layers from the fourth row to the last row in Table \ref{tab:architecture}. The numbers of output channels in Table \ref{tab:architecture} are the specific ones adopted in our experiments and can be set accordingly together with the numbers of intermediate channels. The output dimension sizes are integer-valued in practice. For example, $181\times217\times181$ is downsampled to $91\times109\times91$ via strided convolution, which is upsampled back to $181\times217\times181$ via tri-linear interpolation.

As an overview, VoxHRNet mainly consists of a stem, a backbone and a regression sub-network. The stem sub-network starts with a single strided convolution layer that halves the original resolution as a trade-off between GPU memory usage and loss of shallow information. What follows are two bottleneck modules, each of which includes three convolution layers. The regression sub-network takes as input the concatenation of all levels of feature maps at the end of the backbone sub-network by upsampling, and then passes it through two $1 \times 1 \times 1$ convolution layers, a softmax layer and a tri-linear interpolation layer to produce a soft prediction map.

The backbone sub-network consists of several stages. There are multiple parallel branches corresponding to paths of features maps in different resolutions. Formally speaking, the $k$-th stage retains $k + 1$ branches of feature maps in resolution $1/2, 1/4, \dots, 1/2^{k + 1}$. There is a transitional layer between any two consecutive stages to transform branches in the former stage into those in the latter stage. In order to learn a rich set of features and enlarge receptive field, we place a certain number of high-resolution exchange modules in each stage, with several basic residual blocks in each exchange module, where each block consists of two $3 \times 3 \times 3$ convolution layers, followed by instance normalization \citep{ulyanov2016instance} and ReLU activation. Note that both basic and bottleneck blocks are residual units where the residual connection is either a data path or a consistent path, depending on input and output channels. We impose fusion among features of all the resolutions by inserting a fusion layer between any two consecutive exchange modules. Fusion and transition are realized by identity mapping, $1 \times 1 \times 1$ convolution, tri-linear interpolation or strided convolution accordingly. All branches of feature maps are accumulated to a single level after the last stage. In this way, we encourage our network to make use of features in all levels via fusion and retaining them in several branches.

In comparison to FCN and U-Net, the proposed VoxHRNet keeps high-resolution feature maps without downsampling. It also encourages any kinds of fusion among feature maps from all levels. The final aggregation of multi-scale representations ensures that both global and local information are retained.

This network architecture design is appropriate for whole brain segmentation. Throughout the forward passing process, the network maintains high-resolution representations as well as low-resolution representations by connecting convolutions in the branch for each resolution in parallel. Since we maintain features in all resolution levels, there is no need to recover high-resolution representations via upsampling. To enable feature reuse and learn more reliable features, the network performs multi-scale fusion repeatedly, in which representations of each resolution constantly exchange information with representations of other resolutions. This kind of fusion resembles a fully connected layer where there are all possible connections among feature maps with different sizes and channels. Such a dense fusion strategy also renders it unnecessary to aggregate low-level and high-level features further. After the backbone sub-network, representations of all resolutions are aggregated and fully leveraged for final prediction. \citet{sun2019deep} empirically showed the effectiveness of repeated multi-scale fusion and resolution maintenance, whereas \citet{sun2019high} showed the importance of incorporating low-resolution representations together with high-resolution representations. On that account, VoxHRNet is expected to learn high-resolution representations that are strong enough for whole brain segmentation.

\subsection{Memory-efficient training}

Roughly speaking, an FCN with more parameters has a larger hypothesis space to better fit training data. In spite of that, it is usually not applicable to train an expressive 3D FCN model because its GPU memory usage grows rapidly as the network size increases. Given this consideration, we adopt a reduced-precision DNN training technique called mixed precision training to attain memory-efficient training.

The 16-bit base-2 half-precision floating-point format (FP16) in the IEEE 754-2008 standard \citep{cornea2009ieee} is adopted in this paper in place of the IEEE 754 single-precision floating-point format (FP32). Formally, a floating-point number in FP16 and FP32 in its normal exponent range can be written as

\begin{equation}
    \begin{split}
    V_{FP16} & = (-1)^{(S)_2} \times (1.M)_2 \times 2^{(E)_2-15}, \\
    V_{FP32} & = (-1)^{(S)_2} \times (1.M)_2 \times 2^{(E)_2-127}, \\
    \end{split}
\end{equation}
where $S$ is the sign, $E$ denotes the exponent, $M$ represents the mantissa and $(\cdot)_{radix}$ indicates the base of a numerical representation. The representable positive normal number in FP32 ranges from $1.175 \times 10^{-38}$ to $3.403 \times 10^{38}$ whereas for FP16 it is from $6.104 \times 10^{-5}$ to $6.550 \times 10^4$.

We adopt some approaches as demonstrated by \citet{micikevicius2018mixed} to compensate for loss of information in the half-precision setting. To begin with, a single-precision master copy of each network parameter is maintained for updating. One reason for keeping a master FP32 weight copy is that weight values in FP16 would become zero whenever an FP32 number to be converted to FP16 has an exponent $(E)_2-127 \leq -24$. One possible reason is that the gradient would zero out if the ratio of the weight to the weight gradient is large. Formally, assuming $E_w$ and $E_{\Delta_w}$ are the exponent of the FP16 scalar weight parameter $w$ and its corresponding weight update $\Delta_w$ respectively, the implicit bit in $\Delta_w$ would be shifted strictly more than $11$ bits to the right if $E_w - E_{\Delta_w} \geq 12$. Maintaining single-precision copies resolves this issue and does not add much overhead because activation is the one that dominates memory consumption during training.

Loss scaling is crucial to successful mixed precision training because weight updates with small magnitudes are prevalent during training. As suggested by \citet{micikevicius2018mixed}, approximately $30\%$ of the weight updates in FP32 are not representable with the FP16 format and only about $5\%$ of the gradient values with an exponent value $-24 \leq E \leq -10$ are subsumed by the FP16 representable range. Empirically multiplying the gradient values with an appropriate constant scaling factor would make more gradients covered and thus match the accuracy achieved by normal single-precision training, so long as the gradients are unscaled back between backpropagation and weight clipping. By chain rule, the loss scaling method ensures that all the parameters are scaled up by the same constant. For better adaptability, we incorporate a dynamic loss scaling approach during training. The dynamic loss scale scheduler would increase the current scale if it is valid for a consecutive number of iterations, or decrease it when an overflow issue occurs, where the optimizer skips this weight updating step.

Also, FP16 arithmetic operations are generally faster than FP32 owing to the high computational throughput on generalized matrix-matrix multiplications (GEMMs). Specifically, each Tensor Core performs an FP16 matrix multiplication operation with accumulation in FP32 as follows:

\begin{equation}
    \boldsymbol{D}_* = \boldsymbol{A}_{16} \times \boldsymbol{B}_{16} + \boldsymbol{C}_*,
\end{equation}
where $\boldsymbol{A}_{16}, \boldsymbol{B}_{16} \in \mathbb{R}^{4 \times 4}$ are FP16 matrices and $\boldsymbol{C}_*, \boldsymbol{D}_* \in \mathbb{R}^{4 \times 4}$ are either FP16 or FP32 matrices to accumulate the product results. The fused multiply-add mixed-precision operations are several times faster than standard FP32 operations. Accumulation matrices $\boldsymbol{C}$ and $\boldsymbol{D}$ ensure that there is little arithmetic precision loss. In addition, the arithmetic computational speed is unaffected by carrying out reductions in FP32 because most of the related neural network layers are memory-bandwidth limited. The speed is also insensitive to arithmetic precision of the memory-bandwidth limited point-wise operations.

\subsection{Training data scale increasing}

Data scarcity becomes a severe issue when training DNNs, especially in brain segmentation with the full volume framework. To overcome this and increase training data scale, we propose to incorporate standard volumetric image data augmentation during training by assuming transformation invariance in conditional label distribution. The augmentation methods are required to approximate true distribution and yield low computational cost.

We adopt a volume-level and a spatial-level data augmentation method for whole brain volumes as well as labels. Specifically, for an input volume $X \in \mathbb{R}^{H \times W \times D}$ and its label volume $Y \in \{1, \dots, L\}^{H \times W \times D}$, our goal is to obtain a novel training data sample $X' \in \mathbb{R}^{H' \times W' \times D'}, Y' \in \{1, \dots, L\}^{H' \times W' \times D'}$.

The volume-level augmentation is additive Gaussian noise for the intensity volume,

\begin{equation}
    X' = X + G,
\end{equation}
where $G \sim \mathcal{N}(0, \sigma^2)^{H \times W \times D}$.

The spatial-level augmentation is elastic deformation \citep{simard2003best}, achieved by warping $X$ and $Y$ via a sampling grid generated from a regular grid along with a random displacement field. Formally, in the beginning, for any $v \in \Omega'$, any dimension $J \in \{H', W', D'\}$, a displacement field $d_v^J$ is randomly sampled from a uniform distribution $\mathcal{U}(-1, 1)$ and smoothed as follows:

\begin{equation}
    \begin{split}
    & k_G(x, y, z; \sigma) = \frac{1}{(\sqrt{2\pi}\sigma)^3} e^{-\frac{x^2 + y^2 + z^2}{2\sigma^2}}, \\
    & \Delta_v^J = \alpha^J \sum_{r \in \Omega'} d_{r}^J k_G(x_v - x_r, y_v - y_r, z_v - z_r), \\
    \end{split}
\end{equation}
where $k_G$ means a three-dimensional Gaussian kernel function with $\sigma^2$ being the variance parameter, $(x_r, y_r, z_r)$ signifies the coordinate of voxel $r$, $\Delta_v^J$ stands for the final displacement of voxel $v$ after smoothing, and $\alpha^J$ serves as a scaling factor for computing the actual offset. It is noteworthy that a large $\sigma$ in the Gaussian kernel results in zero displacement field whereas a small $\sigma$ leads to a randomly distributed displacement field. The distorted output feature maps are thus obtained by sampling and interpolating for any voxel $v \in \Omega'$ as follows:

\begin{equation}
    \begin{split}
    & (x_v^s, y_v^s, z_v^s) = (x_v^t, y_v^t, z_v^t) + (\Delta_v^{H'}, \Delta_v^{W'}, \Delta_v^{D'}), \\
    & I_v^t = \sum_{r \in \Omega} I_r^s \cdot \delta(\lfloor x_v^s + 0.5 \rfloor - x_r) \\
    & \cdot \delta(\lfloor y_v^s + 0.5 \rfloor - y_r) \cdot \delta(\lfloor z_v^s + 0.5 \rfloor - z_r), \\
    \end{split}
\end{equation}
where $(x_v^t, y_v^t, z_v^t)$ comes from a regular grid of the output volume, $(x_v^s, y_v^s, z_v^s)$ defines the point-wise correspondence in $X$ for voxel $v$ in the output volume, $I_r^{(\cdot)}$ is the intensity value of voxel $r$, and $\delta(\cdot)$ represents the Kronecker delta function. The above convolution filter corresponds to nearest-neighbor interpolation. Similar elastic deformation with the same displacement field as above is applied to label volume transformation from $Y$ to $Y'$.

\section{Experiments and results}
\label{sec:experimentsandresults}

\subsection{Data preparation}

A public dataset is adopted in the experiments.

The LONI LPBA40 dataset \citep{shattuck2008construction} consists of 40 T1-weighted MRI brain images, from subjects aged $19.3$ to $39.5$, with 54 manually labeled brain ROIs excluding brainstem and cerebellum. Each MRI or label volume is of size $181 \times 217 \times 181$ with $1 \times 1 \times 1$ $mm^3$ voxel size. The dataset is randomly partitioned into $4$ equal-sized subsets for four-fold cross-validation.


\begin{figure}[!t]
    \centering
    \includegraphics[width=0.8\columnwidth]{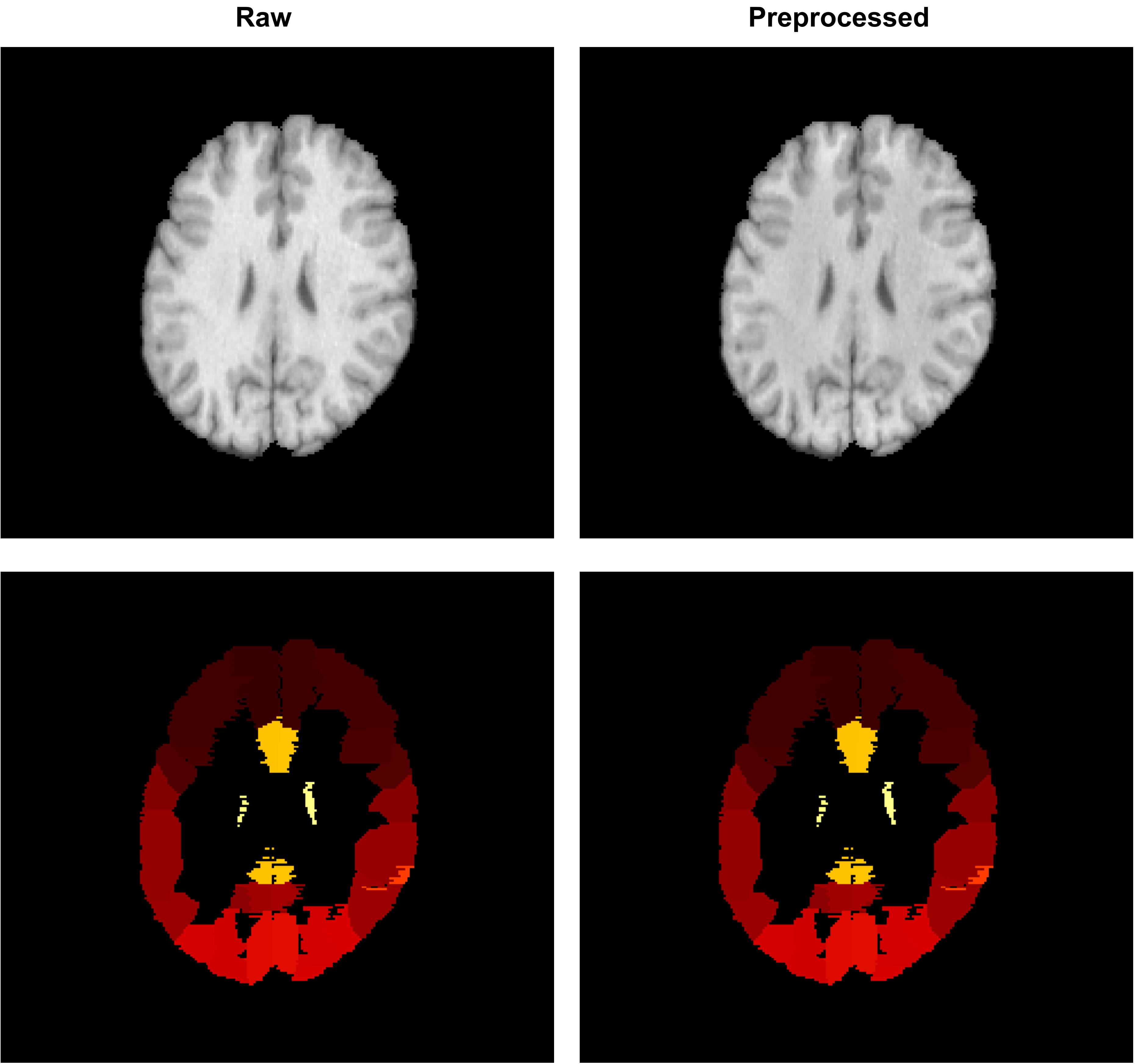}
    \caption{Comparison of an unpreprocessed brain MRI volume and a preprocessed one in the LPBA40 dataset.}
    \label{fig:rawpreprocess}
\end{figure}

N4 bias field correction \citep{tustison2010n4itk} is applied to each volume as the first step. The MRI volumes in the LPBA40 dataset are well-preprocessed, hence left without further correction. Figure \ref{fig:rawpreprocess} presents an example of a pair of unpreprocessed and preprocessed volumes.
The intensity values in each volume are standardized by z-score normalization before forward passing.

\subsection{Implementation details}
\label{sec:impldetail}

We used the implementations of N4 bias field correction and affine registration in the Advanced Normalization Tools (ANTs) package \citep{avants2009advanced} for preprocessing.

Our proposed method is implemented with PyTorch \citep{paszke2019pytorch}. The number of intermediate feature channels in the stem sub-network is set to $32$ and the bottleneck blocks have $64$ channels as output. The number of feature maps retained in each of the $b$ branches from low to high level is set to $16, 32, \dots, 2^{b + 3}$. For the LPBA40 dataset, there are $2$ preset stages in the backbone sub-network, in which the number of exchange modules is set to $1$ and $2$ in each branch for the first and second stage respectively, where each module is composed of $3$ basic residual blocks. We denote the network model for LPBA40 by VoxHRNet.

We impose no bias term in all the convolution layers except for the regression sub-network. Epsilon and momentum in all instance normalization layers are set to $10^{-5}$ and $10^{-1}$ respectively. Weight parameters in network layers are randomly initialized before training. By default, we train our models with the cross-entropy loss. The smoothing constant is set to $0.0001$ for the dice loss. The sigma parameter in elastic transformation is randomly sampled from $\mathcal{U}(0.04, 0,06)$ for each dimension independently whereas sigma in additive Gaussian noise is randomly sampled from $\mathcal{U}(0.0, 0.1)$. To minimize the objective with backpropagation, we adopt RAdam \citep{Liu2020On} as the optimization method with parameters $\beta_1 = 0.9$, $\beta_2 = 0.999$, $\epsilon = 10^{-8}$ and a learning rate of $0.001$, without weight decay. Our model is trained for $30,000$ steps with a batch size of $1$. We adopt an optimization level of $O1$ in the NVIDIA Apex Amp tool as the mixed precision training scheme.

We include FastSurfer \citep{henschel2020fastsurfer} as one of our baseline methods. We use a batch size of $8$ during training and default recommended values for all the other hyperparameters in FastSurfer.

All experiments are conducted on a server with a TITAN Xp 12GB GPU, a desktop with a TITAN RTX 24GB GPU and the same desktop with an Intel(R) Xeon(R) CPU E5-2643 0 @ 3.30GHz CPU. To allow a fair comparison, all the results about GPU memory usage, inference time and number of parameters are re-estimated on the desktop with a 24GB GPU.

\subsection{Evaluation metrics}

Two metrics are adopted to evaluate segmentation results.

The Dice similarity coefficient (DSC), or Dice overlap, is a direct comparison of prediction and ground truth, defined as

\begin{equation}
    DSC(G, P) = \frac{2|G \cap P|}{|G| + |P|} \times 100\%,
\end{equation}
where $G$ stands for the ground truth and $P$ indicates the predicted segmentation volume, with $|\cdot|$ denotes the cardinality of a set of points. A higher DSC implies larger volume similarity thus better segmentation performance. 

The Hausdorff distance (HD) is the maximum Euclidean distance among all the minimum Euclidean distances between two finite points sets, written as

\begin{equation}
    HD(G, P) = \max(\max_{g \in G}\min_{p \in P}\lVert g - p \rVert, \max_{p \in P}\min_{g \in G}\lVert p - g \rVert),
\end{equation}
where $\lVert \cdot \rVert$ is the Euclidean norm. A smaller HD value results from a smaller surface distance between two sets, thus can be regarded as an indicator for better segmentation performance. We compute all the HDs in voxels as adopted in \citet{fang2019automatic}.

We also perform $2$-tailed paired t-test with a small p-value ($p < 0.05$) corresponding to significant difference.

\subsection{LONI LPBA40 dataset}

\subsubsection{Comparison with existing methods}

We perform four-fold cross-validation and compare against baselines including FastSurfer \citep{henschel2020fastsurfer}, HSPBL \citep{wu2015hierarchical}, JLF \citep{wang2013multi}, FCN \citep{long2015fully}, U-Net \citep{ronneberger2015u} and MA-FCN \citep{fang2019automatic}, following the same setup in \citet{fang2019automatic}. Note that HSPBL and JLF are multi-atlas-based methods whereas FCN, U-Net and MA-FCN are patch-based deep learning models. FastSurfer is a slice-based method with three 2D CNNs. Our implementation of the full volume segmentation network instance is described in Section \ref{sec:impldetail}, trained with the cross-entropy loss on the server with a 12GB GPU in mixed precision.

\begin{table}[!t]
    \renewcommand{\arraystretch}{1.3}
    \caption{Comparison of the proposed model with state-of-the-art baselines on the LONI LPBA40 dataset. HSPBL and JLF are multi-atlas segmentation methods whereas FCN, U-Net and MA-FCN are implemented as patch-based methods as in \citet{fang2019automatic}. FastSurfer is slice-based. VoxHRNet includes $1$ and $2$ modules per branch in the first and second stage respectively, where each module has $3$ basic blocks, trained with the cross-entropy loss and mixed precision.
    The results for HSPBL, JLF, FCN, U-Net and MA-FCN are those reported in \citet{fang2019automatic}. The test time of HSPBL and JLF is indicated by a hyphen because it does not apply for multi-atlas-based methods. The test time of other methods is the average forward passing time for one volume. (*) indicates that FastSurfer does not make any prediction for some structures.}
    
    
    \centering
    \resizebox{\columnwidth}{!}{
        \begin{tabular}{cccc}
            \hline
            Method & Test time & DSC (\%) & HD (voxel) \\
            \hline
            FastSurfer & 57 s & 57.33 $\pm$ 11.14 & INF $\pm$ NaN * \\
            HSPBL & - & 78.47 $\pm$ 2.33 & 22.95 $\pm$ 4.81 \\
            JLF & - & 79.19 $\pm$ 0.98 & 17.59 $\pm$ 3.14 \\
            FCN & 90 s & 78.88 $\pm$ 1.07 & 21.50 $\pm$ 4.69 \\
            U-Net & 90 s & 79.42 $\pm$ 1.12 & 16.25 $\pm$ 4.00 \\
            MA-FCN & 140 s & 81.19 $\pm$ 1.06 & 14.11 $\pm$ 3.22 \\
            \hline
            VoxHRNet & \textbf{0.653s} & \textbf{83.12 $\pm$ 3.52} & \textbf{8.89 $\pm$ 2.25} \\
            \hline
        \end{tabular}
    }
    \label{tab:lpba_main}
\end{table}

\begin{figure}[!t]
    \centering
    \includegraphics[width=\columnwidth]{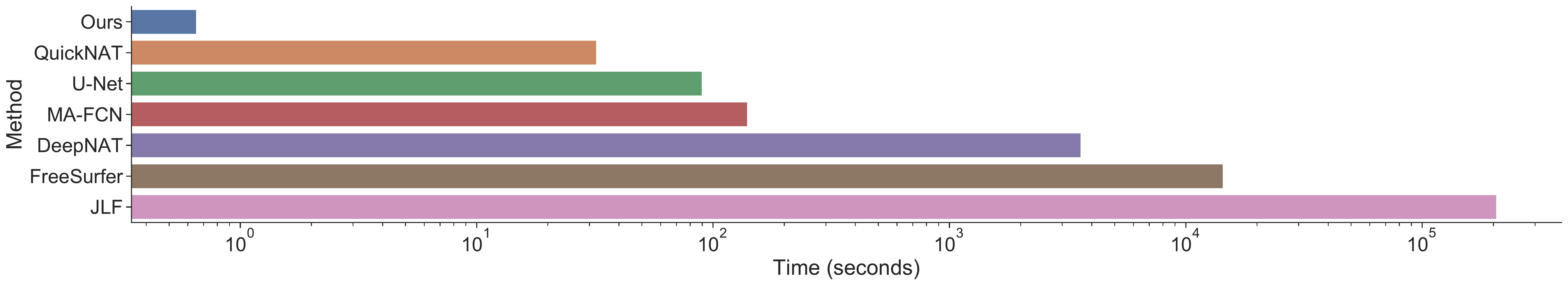}
    \caption{Comparison of approximate time to segment a preprocessed volume in the LPBA40 dataset given $30$ atlas volumes. The only preprocessing step for the LPBA40 dataset is N4 bias field correction. For deep learning methods, the time refers to the total forward passing time. For multi-atlas-based methods, it refers to the sum of registration and label fusion time. The comparisons were carried out with the same data under exactly the same conditions. The time axis is on a logarithmic scale for better exposition. Note that DeepNAT and QuickNAT are able to identify $25$ and $27$ brain structures only.}
    \label{fig:timescale}
\end{figure}

Quantitative results are reported in Table \ref{tab:lpba_main}, where most of the numbers for baselines are from \citet{fang2019automatic} since our experimental settings are identical to theirs. The mean DSC and HD values as well as their standard deviations are computed across all the structures, whereas the average forward passing time is taken as test time. Compared to the traditional multi-atlas methods such as HSPBL \citep{wu2015hierarchical}, JLF \citep{wang2013multi}, our models provide a gain of $4\%$ in terms of mean DSC and a relative improvement of $50\%$ in terms of mean HD. Compared to patch-based methods such as FCN \citep{long2015fully}, U-Net \citep{ronneberger2015u}, MA-FCN \citep{fang2019automatic}, our models improve their best performance by about $2\%$ increase in mean DSC and relatively $32.42\%$ decrease in average HD. We observe that FastSurfer fails in this dataset. Patch-based methods are not comparable to our full-volume-based method possibly because of their drawback that the extracted features are still based on partial data no matter how large the receptive field is.

The inference time excluding preprocessing is on average about $200$ times less than the state-of-the-art baselines owing to our full volume framework. Estimated on the desktop with an Intel(R) Xeon(R) CPU E5-2643, the preprocessing time is about $277$ seconds. Therefore VoxHRNet is about $17\%$ faster than the competitive methods in terms of the overall inference time that includes preprocessing. A vivid illustration comparing test times on a logarithmic scale is shown in Figure \ref{fig:timescale}, where we estimate the total time of segmenting one preprocessed unseen volume in the same environment (the 24G desktop) for methods such as DeepNAT \citep{wachinger2018deepnat}, QuickNAT \citep{roy2019quicknat}, FreeSurfer \citep{fischl2012freesurfer} and JLF \citep{wang2013multi}.

\subsubsection{Ablation study on augmentations}
\label{sec:expaug}

\begin{table}[!t]
    \renewcommand{\arraystretch}{1.3}
    \caption{An ablation study of augmentations on the LPBA40 dataset.}
    
    %
    %
    %
    
    \centering
    \resizebox{\columnwidth}{!}{
        \begin{tabular}{ccc}
            \hline
            Method & DSC (\%) & HD (voxel) \\
            \hline
            No augmentation & 80.57 $\pm$ 3.82 & 10.01 $\pm$ 2.74 \\
            Gaussian & 80.71 $\pm$ 3.83 & 10.66 $\pm$ 3.29 \\
            Elastic distortion & 83.09 $\pm$ 3.61 & 8.92 $\pm$ 2.29 \\
            \hline
            Elastic \& Gaussian & \textbf{83.12 $\pm$ 3.52} & \textbf{8.89 $\pm$ 2.25} \\
            \hline
        \end{tabular}
    }
    \label{tab:augmentation}
\end{table}

Table \ref{tab:augmentation} lists the results under four augmentation settings, namely, no augmentation, random Gaussian noise only, elastic deformation only, and both augmentation methods included, with other settings fixed. Note that the setting with both augmentations corresponds to that described in Section \ref{sec:impldetail}, or equivalently VoxHRNet in Table \ref{tab:lpba_main}. We can conjecture that elastic distortion covers a wide range of brain variations and respects specific transformation invariance property of MRI brain volumes, hence leading to a significant improvement over the setting without elastic distortion. Moreover, we find that the result of both augmentations is not significantly better than that of elastic deformation ($p = 0.45$, $2160$ samples for each method), in which the benefit from deformation may dominate that from Gaussian noise. In contrast, adopting random Gaussian noise achieves significantly better performance in terms of mean DSC than adopting no augmentation ($p = 0.002$, $2160$ samples for each method). Therefore, additive random Gaussian noise is slightly beneficial to final performance because some outlier intensity values produced by the scanner might be accounted for so as to robustify the learned model. To conclude, data augmentation resolves the data scarcity issue and plays an important role in our full volume approach for brain segmentation.

\subsubsection{Ablation study on optimization levels of mixed precision}

\begin{table}[!t]
    \renewcommand{\arraystretch}{1.3}
    \caption{Ablation test results for training techniques with different precisions, evaluated on the LPBA40 dataset. The evaluation results for half precision (O3) are low in DSC and infinite in HD because it diverges.}
    
    %
    %
    %
    
    \centering
    \resizebox{\columnwidth}{!}{
        \begin{tabular}{cccccc}
            \hline
            Method & GPU (MB) & Train time & Test time & DSC (\%) & HD (voxel) \\
            \hline
            Half precision ($O3$) & \textbf{10976} & 34.091h & 0.674s & 2.79 $\pm$ 4.74 & INF $\pm$ NaN \\
            Mixed precision ($O2$) & 10981 & 35.901h & \textbf{0.639s} & 83.03 $\pm$ 3.60 & 8.98 $\pm$ 2.34 \\
            Mixed precision ($O1$) & 11348 & 34.721h & 0.653s & \textbf{83.12 $\pm$ 3.52} & \textbf{8.89 $\pm$ 2.25} \\
            Full precision ($O0$) & 15522 & \textbf{29.585h} & 0.654s & 83.02 $\pm$ 3.57 & 9.07 $\pm$ 2.33 \\
            \hline
        \end{tabular}
    }
    \label{tab:precision}
\end{table}

We investigate the impact of different mixed precision optimization levels on segmentation performance. In the NVIDIA Apex Amp tool, $O0$ corresponds to no precision reduction with all the FP32 operations unchanged. $O1$ is our choice in this work where loss scaling is dynamically applied, and operations that especially benefit from FP32 precision are not cast to FP16. In contrast, $O2$ level converts model weights and input data to FP16, and meanwhile maintains FP32 master copies. $O3$ keeps training entirely in FP16 precision.

We vary the optimization level with other settings fixed, where $O1$ corresponds to VoxHRNet in Table \ref{tab:lpba_main}. This ablation study produces the results in Table \ref{tab:precision}. Optimization level $O0$ demonstrates the least training time but the highest GPU memory usage. Although the half-precision optimization level $O3$ has the minimum GPU memory requirement, we fail to train our model in this setting, where the computed gradients are unstable as a result of the vulnerability of FP16 to overflow. As precision together with the number of FP32 operations decreases, the overall GPU memory usage monotonically decreases in that the number of bytes accessed is directly reduced. The difference between the training times of $O0$ and those levels involving FP16 depends on GPU support and specific implementation of mixed precision training, where we observe a slight increase in training time if precision is reduced and floating-point conversion is necessary. Among optimization levels $O1$-$O3$, $O3$ has the least training time and usually serves as a speed baseline for $O1$ and $O2$. Despite the fact that training fails with $O3$, less aggressive precision levels including $O1$ and $O2$ introduce a little stochasticity and regularization to the gradients, which are shown to benefit backpropagation learning empirically. To sum up, $O1$ gives a good trade-off choice with the best segmentation performance and a significant memory reduction compared to fully FP32 precision training.

Given the network architecture in Section \ref{sec:impldetail}, we train it with isotropic brain volumes and find the maximum deployable sizes in the 12GB GPU server for $O0$ and $O1$ respectively. We observe an increase of about $40\%$ in the maximum deployable volume size from $175 \times 175 \times 175$ to $194 \times 194 \times 194$. The memory reduction benefits brought by mixed precision training are not as good as what are shown theoretically because factors including GPU memory consumption of the network parameters, memory bandwidth of the layers and generated computational graphs have to be taken into account in practice. Furthermore, after applying rescaling to the original resolution, the mean dice result of the network trained with $O0$ level and volumes of size $175^3$ is $0.7875$ while it is $0.7942$ for the network trained with $O1$ level and volumes of size $194^3$, where the corresponding mean HD values are $11.83$ and $9.96$ respectively. This indicates that being able to train with volumes in higher resolution can possibly improve model's generalizability.

\subsubsection{Ablation study on network architectures and sizes}

\begin{table}[!t]
    \renewcommand{\arraystretch}{1.3}
    \caption{Performance comparison of our proposed FCN with state-of-the-art networks in the full volume framework, evaluated on the LONI LPBA40 dataset. For fair comparison, U-Net, VoxResNet and U-Net++ contain a similar number of parameters to VoxHRNet. VoxHRNet (small) includes $12$, $24$ and $48$ feature maps from low to high level branches. VoxHRNet includes $16$, $32$ and $64$ feature maps. VoxHRNet (large) includes $24$, $48$ and $96$ feature maps.}
    
    
    \centering
    \resizebox{\columnwidth}{!}{
        \begin{tabular}{ccccc}
            \hline
            Method & \# params & Test time & DSC (\%) & HD (voxel) \\
            \hline
            U-Net (full volume) & 2.1M & 0.682s & 80.23 $\pm$ 3.75 & 34.23 $\pm$ 10.57 \\
            VoxResNet (full volume) & 1.9M & 0.881s & 82.29 $\pm$ 3.70 & 10.13 $\pm$ 2.53 \\
            U-Net++ (full volume) & 2.2M & 1.568s & 81.39 $\pm$ 3.95 & 13.91 $\pm$ 4.78 \\
            \hline
            VoxHRNet (small) & \textbf{1.4M} & \textbf{0.576s} & 82.84 $\pm$ 3.58 & 9.02 $\pm$ 2.21 \\
            VoxHRNet & 2.4M & 0.653s & 83.12 $\pm$ 3.52 & \textbf{8.89 $\pm$ 2.25} \\
            VoxHRNet (large) & 5.1M & 0.736s & \textbf{83.18 $\pm$ 3.56} & 8.99 $\pm$ 2.36 \\
            \hline
        \end{tabular}
    }
    \label{tab:lpba40_paras}
\end{table}

\begin{figure*}[!t]
    \centering
    \includegraphics[width=\textwidth]{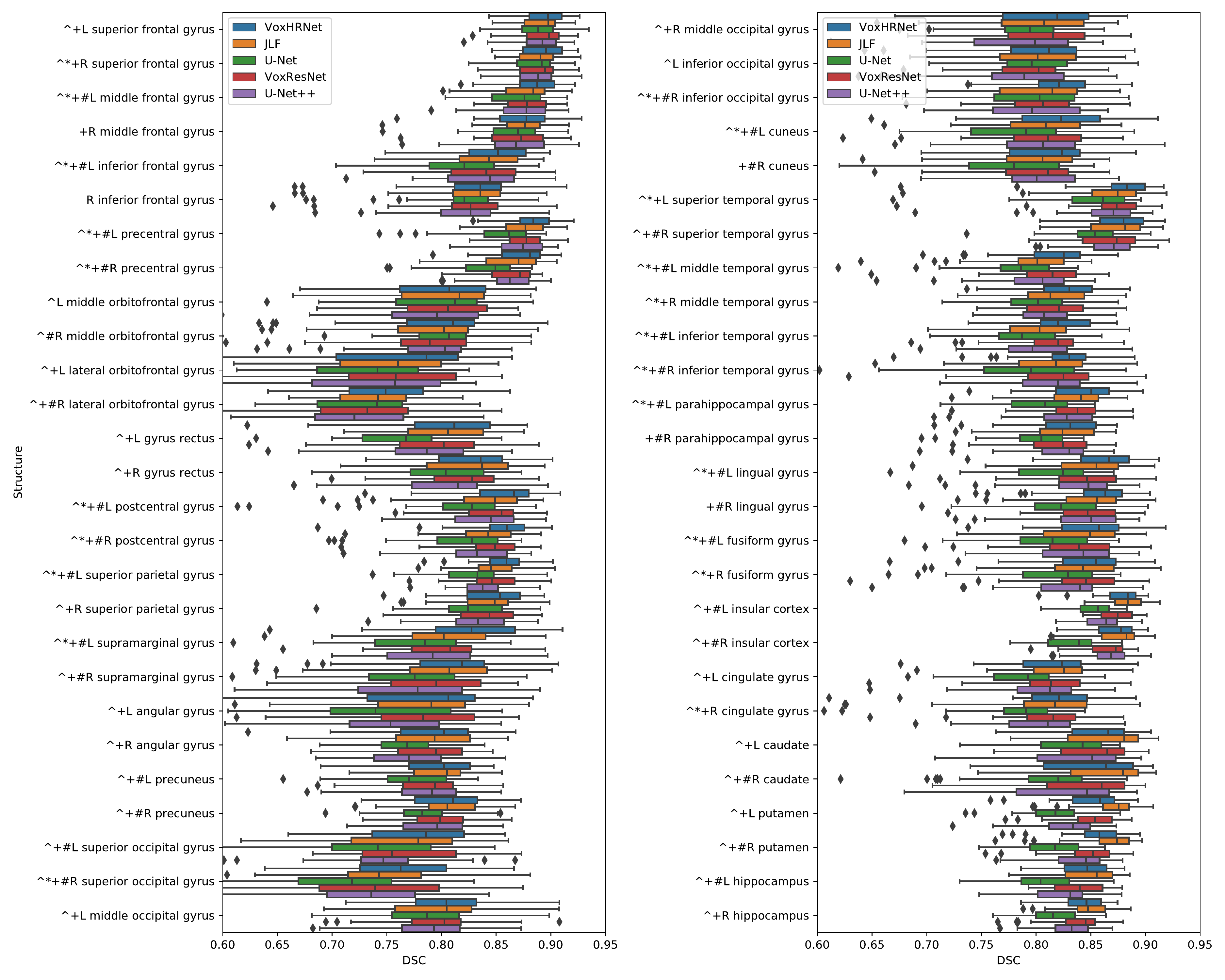}
    \caption{Box plot of structure-wise DSC comparing our method with JLF, U-Net, VoxResNet and U-Net++ on the LONI LPBA40 dataset. The symbols '*', '+', '\#' and '\^{}' indicate statistically significant improvements ($p < 0.05$) of our method in terms of DSC over JLF, U-Net, VoxResNet and U-Net++ respectively. Best viewed in color.}
    \label{fig:lpba40classwise}
\end{figure*}

\begin{figure}[!t]
    \centering
    \includegraphics[width=\columnwidth]{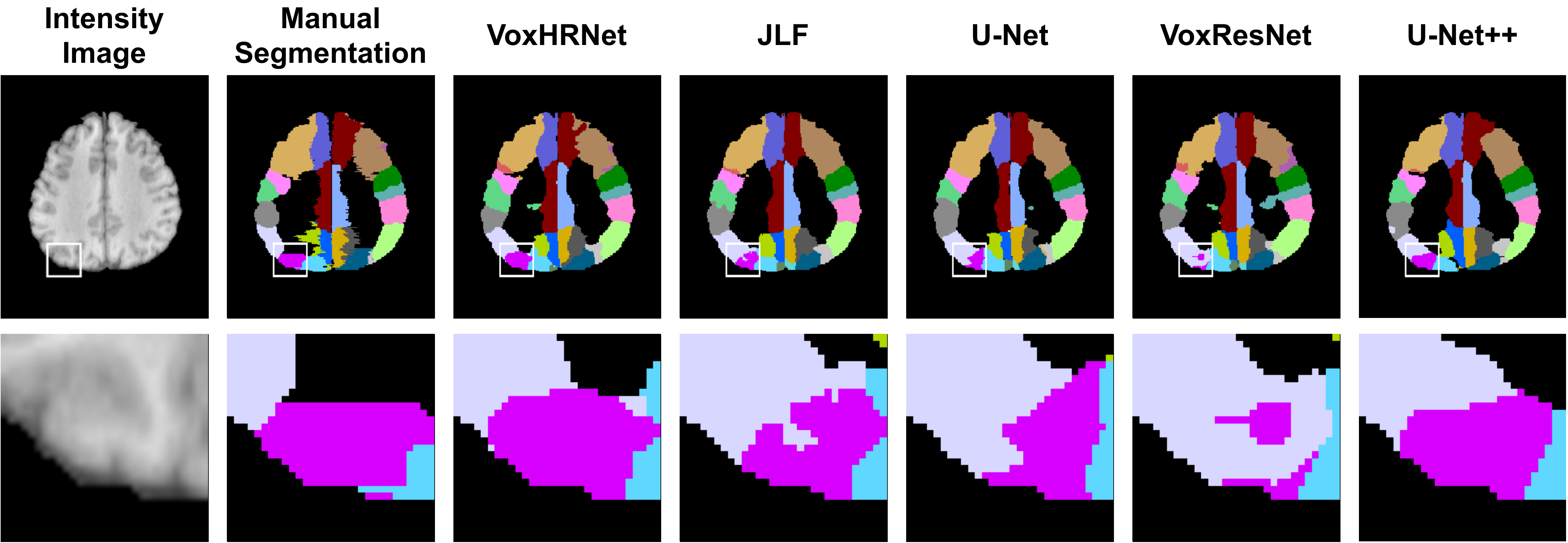}
    \caption{Visual comparison of segmentation results of our method with JLF, U-Net, VoxResNet and U-Net++ on the LONI LPBA40 dataset. A slice in a typical MRI volume is selected to illustrate for each dataset. A region is randomly chosen and zoomed in below the original image for better exposition. Each brain structure corresponds to a unique color. A good segmentation result has similar visualization to that of manual segmentation. In the second row, left angular gyrus, left superior occipital gyrus and left middle occipital gyrus are shown in lavender blue, steel blue and magenta respectively. Best viewed in color.}
    \label{fig:segvisual}
\end{figure}

\begin{figure}[!t]
    \centering
    \includegraphics[width=\columnwidth]{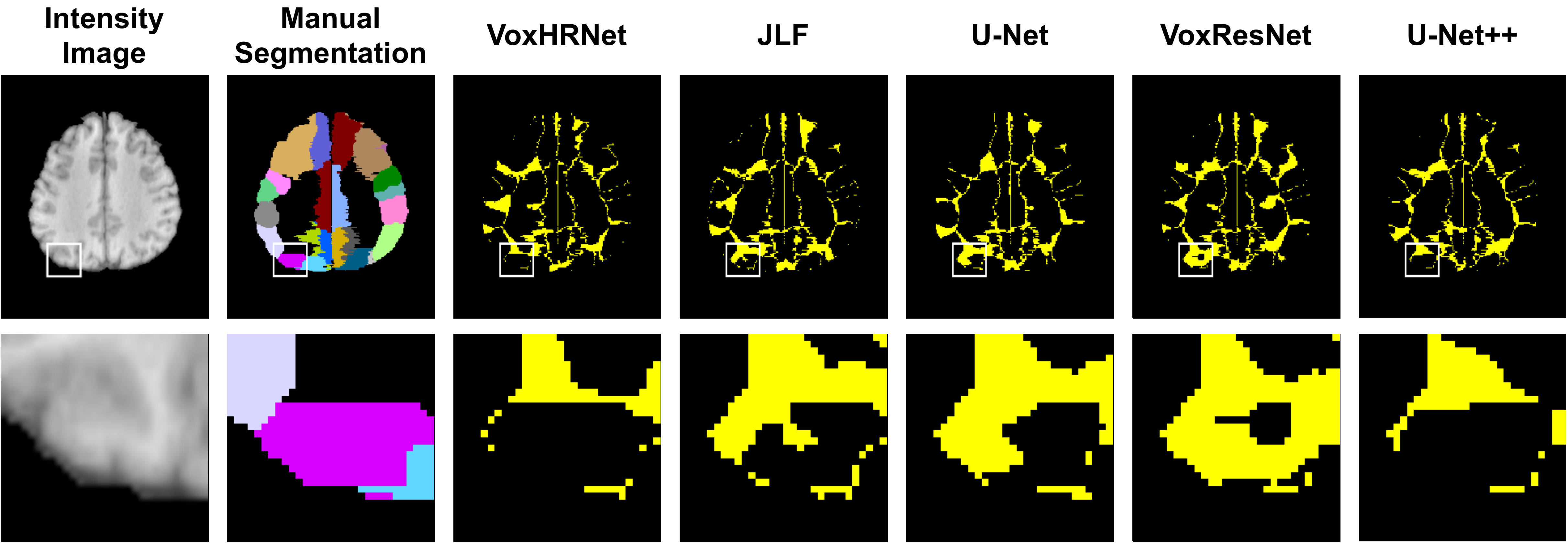}
    \caption{Visual comparisons of false prediction maps of our method with JLF, U-Net, VoxResNet and U-Net++ on the LONI LPBA40 dataset. A slice in a typical MRI volume is selected as illustration for each dataset. A region is randomly chosen and zoomed in below the original image for better exposition. Each brain structure corresponds to a unique color. A mismatch between prediction and ground truth in a voxel is marked in yellow. A good segmentation result has small yellow area in its false prediction map. Best viewed in color.}
    \label{fig:fpvisual}
\end{figure}

To gain more insights about the adopted network architecture, we reimplement U-Net, VoxResNet and U-Net++ \citep{zhou2019unetplusplus, zhou2018unetplusplus} by adapting them to the full volume framework with other settings fixed. In other words, we replace only the network structure in our framework with U-Net, VoxResNet or U-Net++ accordingly. Note that the same stem sub-network of VoxHRNet in Table \ref{tab:lpba_main} is placed at the beginning of forward passing in VoxResNet and U-Net++ because it generally fails to be deployed in a 12G GPU platform. U-Net in the full volume framework can be trained successfully because we adaptively decrease its feature channels and adopt mixed precision training. Our model is indicated by VoxHRNet, identical to VoxHRNet in Table \ref{tab:lpba_main}, which shares approximately the same number of parameters with U-Net, VoxResNet and U-Net++. The results are shown in Table \ref{tab:lpba40_paras} and a detailed structure-wise DSC comparison is illustrated in Figure \ref{fig:lpba40classwise} with significance test. With a comparable network size to that of the full volume variants U-Net, VoxResNet and U-Net++, VoxHRNet increases mean DSC by about $0.6\%$ and relatively decreases mean HD by about $12\%$. Significance test results ($2160$ samples for each method) validate that we are statistically significantly better ($p < 0.05$) than all the baselines. We obtain qualitative results by visualizing the segmentation of a slice within a typical volume from the LPBA40 dataset, as shown in the upper part of Figure \ref{fig:segvisual}. Our predicted segmentation for the left middle occipital gyrus (shown in magenta color centered on the zoomed-in pictures) appears to be more accurate. Similar advantages in segmentation details are observed for other structures and boundaries, which is non-trivial especially in intensity-homogeneous regions. A more detailed illustration of the corresponding voxel-wise false prediction maps is shown in Figure \ref{fig:fpvisual} to corroborate empirical proof of our segmentation performance.

We examine the effect of network size on segmentation performance. To this end, a smaller and larger version of our model are implemented by modifying the network architecture. More specifically, VoxHRNet (small) maintains three branches with $12$, $24$ and $48$ feature channels whereas VoxHRNet (large) encompasses three branches with $24$, $48$ and $96$ feature channels, with other settings identical to that of VoxHRNet. The results for the LPBA40 dataset are reported in Table \ref{tab:lpba40_paras}. A slightly higher segmentation performance can be achieved by our model with much fewer parameters than the baselines. Further increasing the network size results in a performance gain by a small margin at the cost of higher GPU resource demand. As a consequence, a modestly sized VoxHRNet gives almost the best performance, relatively faster inference time and practically applicable GPU utilization.

\begin{figure}[!t]
    \centering
    \includegraphics[width=\columnwidth]{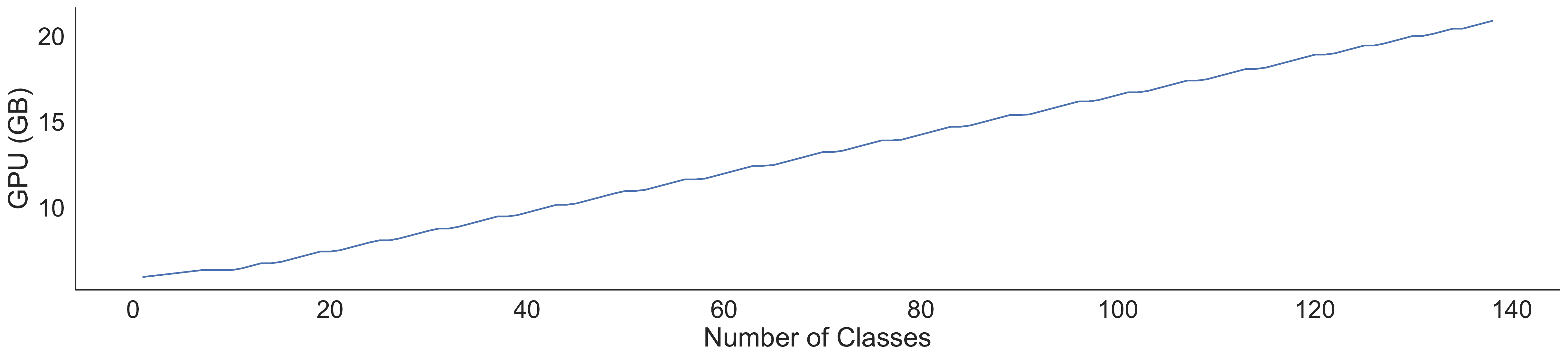}
    \caption{Plot of maximum allocated GPU memory versus the number of structures to be segmented.}
    \label{fig:gpunumclasses}
\end{figure}

Additionally, we plot the maximum allocated GPU memory during training in relation to the number of classes of interest based on VoxHRNet on the LPBA40 dataset, depicted in Figure \ref{fig:gpunumclasses}. The graph demonstrates a roughly linear relationship between them. Hence we believe that a growing amount of GPU memory requirement for volumetric semantic segmentation with more classes cannot be easily avoided with an FCN backbone.

\subsubsection{Ablation study on loss functions}

\begin{table}[!t]
    \renewcommand{\arraystretch}{1.3}
    \caption{Evaluation of segmentation results for different loss functions on the LPBA40 dataset.}
    
    %
    %
    
    \centering
    \resizebox{\columnwidth}{!}{
        \begin{tabular}{cccc}
            \hline
            Method & GPU (MB) & DSC (\%) & HD (voxel) \\
            \hline
            Cross-entropy loss & \textbf{11348} & 83.12 $\pm$ 3.52 & \textbf{8.89 $\pm$ 2.25} \\
            Dice loss & 12913 & 83.09 $\pm$ 3.62 & 24.60 $\pm$ 9.39 \\
            Combined loss & 14477 & \textbf{83.18 $\pm$ 3.57} & 10.27 $\pm$ 2.78 \\
            \hline
        \end{tabular}
    }
    \label{tab:loss}
\end{table}

On top of the cross-entropy loss adopted in most of our experiments, we study the effectiveness of other loss functions as introduced in Section \ref{sec:formulation}. We vary loss function with other settings fixed. The model trained with the cross-entropy loss is the same as VoxHRNet in Table \ref{tab:lpba_main}. As indicated by Table \ref{tab:loss}, the combined loss achieves the best performance regarding DSC whereas the cross-entropy loss achieves the lowest HD. Despite that, the results suggest that any loss that incorporates the dice loss has inferior HD. Intuitively, optimizing a dice loss makes the training objective align with the DSC evaluation metric. However, the dice loss is based on basic cardinalities of the confusion matrix, treating false positives and false negatives equally important given fixed true positives, regardless of surface distance information. Therefore, the gradient computed based on the dice loss may be noisy, resulting in training instability, especially when the dataset involves imbalanced and a large number of classes \citep{kervadec2019boundary}. On the other hand, the cross-entropy loss is a simple regional loss based on information theoretic property from the predicted conditional label distribution, which typically leads to smooth gradients and an implicitly regularized objective. This might explain why the cross-entropy loss is superior to the dice loss in terms of HD. Above all, cross-entropy is a more preferable choice given its excellent performance as to both metrics, as well as its low GPU memory cost.

\section{Discussion and conclusion}
\label{sec:conclusion}

In this paper, we proposed to adopt a full volume learning framework for 3D MRI whole brain segmentation. An effective instantiated model under this framework was given subsequently. To tackle the challenge of training DNNs with large volumetric data, an FCN capable of learning rich features and intricate fusion strategies among network branches was adopted. A mixed precision training technique was incorporated to further facilitate GPU memory efficiency and matrix computation acceleration. Standard augmentation methods including elastic distortion along with additive Gaussian noise were applied to training volumes to address the issue of a very limited amount of available data. The experiment results on the LONI LPBA40 dataset showed that our proposed method achieved state-of-the-art segmentation performance in terms of DSC and HD. The inference time (excluding preprocessing) of our VoxHRNet model was less than $1$ second for segmenting a whole brain volume. Ablation test experiments validated its effectiveness.

The full volume framework is simple and general, with flexibility to integrate different network architectures, reduced-precision training techniques and augmentations. It is superior to the multi-atlas framework because of its significantly faster inference time as well as the end-to-end training scheme. Its advantages over patch-based, slice-based and sub-volume-based approaches lie in simplicity and effectiveness besides the fast inference time. It also outperforms the full volume variants of the competitive network architectures in whole brain segmentation, particularly in segmenting small neuroanatomical regions. Apart from the advantages of VoxHRNet, our full volume framework considerably reduces the GPU memory requirement via mixed precision training, further makes many more neural network models able to be incorporated. To summarize, this work demonstrated a straightforward and powerful framework for the challenging whole brain segmentation task.

In addition to augmentation, early stopping was adopted to prevent overfitting. That being said, the best way of minimizing the estimation error in machine learning is usually to acquire more data sampled from the underlying distribution, for example, adopting data with automatically generated labels \citep{huo20193d}. We trained our models with a batch size of $1$ in the experiments because of limited GPU memory and large size of the volumetric data, which may result in more uncertainty in gradient computation and the overall learning process. A more advanced variance reduction optimization algorithm can be considered \citep{zhou2018stochastic}. On the flip side, FCNs with a batch size of $1$ enable us to make use of volumetric data with arbitrary shapes directly without preprocessing steps such as cropping, given sufficient GPU memory. During training, the training risk was monotonically decreasing with a little oscillation because of augmentation and the batch size as expected. On the other hand, we computed the test risk and observed a double descent behavior, studied in \citet{Nakkiran2020Deep}. One explanation is the large hypothesis space of neural network models. When adopting the full volume framework in practice, a noteworthy issue is that a large input volume may cause GPU running out of memory. For that reason, pre-specifying the maximum allowed size of the input volumes for training and testing is necessary if downsampling is undesirable.
Additionally, a hyperparameter search is usually performed whenever we change the dataset, overall network architecture, GPU platform or reduced-precision training scheme. One solution is to automate the process by incorporating techniques of neural architecture search.

There is still space for exploration in improving and applying our proposed method. More effective techniques in reduced-precision training of DNNs can be incorporated to enlarge hypothesis space as well as numerical representable scale. Advances in FCNs and other deep network architectures for semantic segmentation may be beneficial to our method as well. Moreover, applying our framework to multi-modal brain segmentation with T2-weighted volumes and super-resolution segmentation with 7T brain MRI volumes are worth investigation. For example, to tackle multi-modal MRI data, we can train on concatenation of multi-modal volumes, train one network for each modality and perform label fusion in post-processing, or adopt a multi-way stem sub-network to learn high-resolution representation for each modality in the beginning. We discussed inference time that excludes preprocessing in the experiments. In order for clinical use, unpreprocessed data should be considered directly in both training and testing, which we leave to future work. Since the proposed training scheme in this work is quite general, it would henceforth be of interest to study its applicability to various 3D medical image analysis and computer vision tasks.

\section*{Acknowledgments}

This work was supported by the National Science and Technology Major Project of the Ministry of Science and Technology in China under Grant 2017YFC0110903, Microsoft Research under the eHealth program, the National Natural Science Foundation in China under Grant 81771910, the 111 Project in China under Grant B13003.

The authors would like to thank Hongzhi Wang, Shilin Liu, Yang Liu, Longwei Fang and Yuankai Huo for all the helpful discussions.







\bibliography{main}

\begin{thebibliography}{116}
\providecommand{\natexlab}[1]{#1}
\providecommand{\url}[1]{\texttt{#1}}
\providecommand{\urlprefix}{URL }
\expandafter\ifx\csname urlstyle\endcsname\relax
  \providecommand{\doi}[1]{doi:\discretionary{}{}{}#1}\else
  \providecommand{\doi}[1]{doi:\discretionary{}{}{}\begingroup
  \urlstyle{rm}\url{#1}\endgroup}\fi
\providecommand{\bibinfo}[2]{#2}

\bibitem[{Gonz{\'a}lez-Vill{\`a} et~al.(2016)Gonz{\'a}lez-Vill{\`a}, Oliver,
  Valverde, Wang, Zwiggelaar, and Llad{\'o}}]{gonzalez2016review}
\bibinfo{author}{S.~Gonz{\'a}lez-Vill{\`a}}, \bibinfo{author}{A.~Oliver},
  \bibinfo{author}{S.~Valverde}, \bibinfo{author}{L.~Wang},
  \bibinfo{author}{R.~Zwiggelaar}, \bibinfo{author}{X.~Llad{\'o}},
  \bibinfo{title}{A review on brain structures segmentation in magnetic
  resonance imaging}, \bibinfo{journal}{Artificial Intelligence in Medicine}
  \bibinfo{volume}{73} (\bibinfo{year}{2016}) \bibinfo{pages}{45--69}.

\bibitem[{Huo et~al.(2019)Huo, Xu, Xiong, Aboud, Parvathaneni, Bao, Bermudez,
  Resnick, Cutting, and Landman}]{huo20193d}
\bibinfo{author}{Y.~Huo}, \bibinfo{author}{Z.~Xu}, \bibinfo{author}{Y.~Xiong},
  \bibinfo{author}{K.~Aboud}, \bibinfo{author}{P.~Parvathaneni},
  \bibinfo{author}{S.~Bao}, \bibinfo{author}{C.~Bermudez},
  \bibinfo{author}{S.~M. Resnick}, \bibinfo{author}{L.~E. Cutting},
  \bibinfo{author}{B.~A. Landman}, \bibinfo{title}{3d whole brain segmentation
  using spatially localized atlas network tiles}, \bibinfo{journal}{NeuroImage}
  \bibinfo{volume}{194} (\bibinfo{year}{2019}) \bibinfo{pages}{105--119}.

\bibitem[{Fischl(2012)}]{fischl2012freesurfer}
\bibinfo{author}{B.~Fischl}, \bibinfo{title}{FreeSurfer},
  \bibinfo{journal}{NeuroImage} \bibinfo{volume}{62}~(\bibinfo{number}{2})
  (\bibinfo{year}{2012}) \bibinfo{pages}{774--781}.

\bibitem[{Shattuck and Leahy(2002)}]{shattuck2002brainsuite}
\bibinfo{author}{D.~W. Shattuck}, \bibinfo{author}{R.~M. Leahy},
  \bibinfo{title}{BrainSuite: an automated cortical surface identification
  tool}, \bibinfo{journal}{Medical Image Analysis}
  \bibinfo{volume}{6}~(\bibinfo{number}{2}) (\bibinfo{year}{2002})
  \bibinfo{pages}{129--142}.

\bibitem[{Wang and Yushkevich(2013)}]{wang2013multi}
\bibinfo{author}{H.~Wang}, \bibinfo{author}{P.~Yushkevich},
  \bibinfo{title}{Multi-atlas segmentation with joint label fusion and
  corrective learning—an open source implementation},
  \bibinfo{journal}{Frontiers in Neuroinformatics} \bibinfo{volume}{7}
  (\bibinfo{year}{2013}) \bibinfo{pages}{27}.

\bibitem[{Asman and Landman(2013)}]{asman2013non}
\bibinfo{author}{A.~J. Asman}, \bibinfo{author}{B.~A. Landman},
  \bibinfo{title}{Non-local statistical label fusion for multi-atlas
  segmentation}, \bibinfo{journal}{Medical Image Analysis}
  \bibinfo{volume}{17}~(\bibinfo{number}{2}) (\bibinfo{year}{2013})
  \bibinfo{pages}{194--208}.

\bibitem[{Zhang et~al.(2001)Zhang, Brady, and Smith}]{zhang2001segmentation}
\bibinfo{author}{Y.~Zhang}, \bibinfo{author}{M.~Brady},
  \bibinfo{author}{S.~Smith}, \bibinfo{title}{Segmentation of brain MR images
  through a hidden Markov random field model and the expectation-maximization
  algorithm}, \bibinfo{journal}{IEEE Transactions on Medical Imaging}
  \bibinfo{volume}{20}~(\bibinfo{number}{1}) (\bibinfo{year}{2001})
  \bibinfo{pages}{45--57}.

\bibitem[{Tu and Bai(2009)}]{tu2009auto}
\bibinfo{author}{Z.~Tu}, \bibinfo{author}{X.~Bai}, \bibinfo{title}{Auto-context
  and its application to high-level vision tasks and 3d brain image
  segmentation}, \bibinfo{journal}{IEEE Transactions on Pattern Analysis and
  Machine Intelligence} \bibinfo{volume}{32}~(\bibinfo{number}{10})
  (\bibinfo{year}{2009}) \bibinfo{pages}{1744--1757}.

\bibitem[{Krizhevsky et~al.(2012)Krizhevsky, Sutskever, and
  Hinton}]{krizhevsky2012imagenet}
\bibinfo{author}{A.~Krizhevsky}, \bibinfo{author}{I.~Sutskever},
  \bibinfo{author}{G.~E. Hinton}, \bibinfo{title}{Imagenet classification with
  deep convolutional neural networks}, in: \bibinfo{booktitle}{Advances in
  Neural Information Processing Systems}, \bibinfo{pages}{1097--1105},
  \bibinfo{year}{2012}.

\bibitem[{Long et~al.(2015)Long, Shelhamer, and Darrell}]{long2015fully}
\bibinfo{author}{J.~Long}, \bibinfo{author}{E.~Shelhamer},
  \bibinfo{author}{T.~Darrell}, \bibinfo{title}{Fully convolutional networks
  for semantic segmentation}, in: \bibinfo{booktitle}{Proceedings of the IEEE
  Conference on Computer Vision and Pattern Recognition},
  \bibinfo{pages}{3431--3440}, \bibinfo{year}{2015}.

\bibitem[{Fang et~al.(2019)Fang, Zhang, Nie, Cao, Rekik, Lee, He, and
  Shen}]{fang2019automatic}
\bibinfo{author}{L.~Fang}, \bibinfo{author}{L.~Zhang},
  \bibinfo{author}{D.~Nie}, \bibinfo{author}{X.~Cao},
  \bibinfo{author}{I.~Rekik}, \bibinfo{author}{S.-W. Lee},
  \bibinfo{author}{H.~He}, \bibinfo{author}{D.~Shen}, \bibinfo{title}{Automatic
  brain labeling via multi-atlas guided fully convolutional networks},
  \bibinfo{journal}{Medical Image Analysis} \bibinfo{volume}{51}
  (\bibinfo{year}{2019}) \bibinfo{pages}{157--168}.

\bibitem[{Chen et~al.(2018{\natexlab{a}})Chen, Bentley, Mori, Misawa, Fujiwara,
  and Rueckert}]{chen2018drinet}
\bibinfo{author}{L.~Chen}, \bibinfo{author}{P.~Bentley},
  \bibinfo{author}{K.~Mori}, \bibinfo{author}{K.~Misawa},
  \bibinfo{author}{M.~Fujiwara}, \bibinfo{author}{D.~Rueckert},
  \bibinfo{title}{Drinet for medical image segmentation},
  \bibinfo{journal}{IEEE Transactions on Medical Imaging}
  \bibinfo{volume}{37}~(\bibinfo{number}{11})
  (\bibinfo{year}{2018}{\natexlab{a}}) \bibinfo{pages}{2453--2462}.

\bibitem[{Roy et~al.(2019)Roy, Conjeti, Navab, Wachinger, Initiative
  et~al.}]{roy2019quicknat}
\bibinfo{author}{A.~G. Roy}, \bibinfo{author}{S.~Conjeti},
  \bibinfo{author}{N.~Navab}, \bibinfo{author}{C.~Wachinger},
  \bibinfo{author}{A.~D.~N. Initiative}, et~al., \bibinfo{title}{QuickNAT: A
  fully convolutional network for quick and accurate segmentation of
  neuroanatomy}, \bibinfo{journal}{NeuroImage} \bibinfo{volume}{186}
  (\bibinfo{year}{2019}) \bibinfo{pages}{713--727}.

\bibitem[{Sun et~al.(2019{\natexlab{a}})Sun, Ma, Ding, Huang, Liang, and
  Paisley}]{sun20193d}
\bibinfo{author}{L.~Sun}, \bibinfo{author}{W.~Ma}, \bibinfo{author}{X.~Ding},
  \bibinfo{author}{Y.~Huang}, \bibinfo{author}{D.~Liang},
  \bibinfo{author}{J.~Paisley}, \bibinfo{title}{A 3D Spatially Weighted Network
  for Segmentation of Brain Tissue From MRI}, \bibinfo{journal}{IEEE
  Transactions on Medical Imaging} \bibinfo{volume}{39}~(\bibinfo{number}{4})
  (\bibinfo{year}{2019}{\natexlab{a}}) \bibinfo{pages}{898--909}.

\bibitem[{Henschel et~al.(2020)Henschel, Conjeti, Estrada, Diers, Fischl, and
  Reuter}]{henschel2020fastsurfer}
\bibinfo{author}{L.~Henschel}, \bibinfo{author}{S.~Conjeti},
  \bibinfo{author}{S.~Estrada}, \bibinfo{author}{K.~Diers},
  \bibinfo{author}{B.~Fischl}, \bibinfo{author}{M.~Reuter},
  \bibinfo{title}{FastSurfer-A fast and accurate deep learning based
  neuroimaging pipeline}, \bibinfo{journal}{NeuroImage} \bibinfo{volume}{219}
  (\bibinfo{year}{2020}) \bibinfo{pages}{117012}.

\bibitem[{Coup{\'e} et~al.(2020)Coup{\'e}, Mansencal, Cl{\'e}ment, Giraud,
  de~Senneville, Ta, Lepetit, and Manjon}]{coupe2020assemblynet}
\bibinfo{author}{P.~Coup{\'e}}, \bibinfo{author}{B.~Mansencal},
  \bibinfo{author}{M.~Cl{\'e}ment}, \bibinfo{author}{R.~Giraud},
  \bibinfo{author}{B.~D. de~Senneville}, \bibinfo{author}{V.-T. Ta},
  \bibinfo{author}{V.~Lepetit}, \bibinfo{author}{J.~V. Manjon},
  \bibinfo{title}{AssemblyNet: A large ensemble of CNNs for 3D Whole Brain MRI
  Segmentation}, \bibinfo{journal}{NeuroImage} \bibinfo{volume}{219}
  (\bibinfo{year}{2020}) \bibinfo{pages}{117026}.

\bibitem[{Li et~al.(2017)Li, Wang, Fidon, Ourselin, Cardoso, and
  Vercauteren}]{li2017compactness}
\bibinfo{author}{W.~Li}, \bibinfo{author}{G.~Wang}, \bibinfo{author}{L.~Fidon},
  \bibinfo{author}{S.~Ourselin}, \bibinfo{author}{M.~J. Cardoso},
  \bibinfo{author}{T.~Vercauteren}, \bibinfo{title}{On the compactness,
  efficiency, and representation of 3D convolutional networks: brain
  parcellation as a pretext task}, in: \bibinfo{booktitle}{International
  Conference on Information Processing in Medical Imaging},
  \bibinfo{pages}{348--360}, \bibinfo{year}{2017}.

\bibitem[{Ronneberger et~al.(2015)Ronneberger, Fischer, and
  Brox}]{ronneberger2015u}
\bibinfo{author}{O.~Ronneberger}, \bibinfo{author}{P.~Fischer},
  \bibinfo{author}{T.~Brox}, \bibinfo{title}{U-net: Convolutional networks for
  biomedical image segmentation}, in: \bibinfo{booktitle}{International
  Conference on Medical Image Computing and Computer-Assisted Intervention},
  \bibinfo{pages}{234--241}, \bibinfo{year}{2015}.

\bibitem[{Dora et~al.(2017)Dora, Agrawal, Panda, and Abraham}]{dora2017state}
\bibinfo{author}{L.~Dora}, \bibinfo{author}{S.~Agrawal},
  \bibinfo{author}{R.~Panda}, \bibinfo{author}{A.~Abraham},
  \bibinfo{title}{State-of-the-art methods for brain tissue segmentation: A
  review}, \bibinfo{journal}{IEEE Reviews in Biomedical Engineering}
  \bibinfo{volume}{10} (\bibinfo{year}{2017}) \bibinfo{pages}{235--249}.

\bibitem[{Valverde et~al.(2015)Valverde, Oliver, Cabezas, Roura, and
  Llad{\'o}}]{valverde2015comparison}
\bibinfo{author}{S.~Valverde}, \bibinfo{author}{A.~Oliver},
  \bibinfo{author}{M.~Cabezas}, \bibinfo{author}{E.~Roura},
  \bibinfo{author}{X.~Llad{\'o}}, \bibinfo{title}{Comparison of 10 brain tissue
  segmentation methods using revisited IBSR annotations},
  \bibinfo{journal}{Journal of Magnetic Resonance Imaging}
  \bibinfo{volume}{41}~(\bibinfo{number}{1}) (\bibinfo{year}{2015})
  \bibinfo{pages}{93--101}.

\bibitem[{Liew and Yan(2006)}]{liew2006current}
\bibinfo{author}{A.~W.-C. Liew}, \bibinfo{author}{H.~Yan},
  \bibinfo{title}{Current methods in the automatic tissue segmentation of 3D
  magnetic resonance brain images}, \bibinfo{journal}{Current Medical Imaging}
  \bibinfo{volume}{2}~(\bibinfo{number}{1}) (\bibinfo{year}{2006})
  \bibinfo{pages}{91--103}.

\bibitem[{Xue et~al.(2001)Xue, Ruan, Moretti, Revenu, and
  Bloyet}]{xue2001knowledge}
\bibinfo{author}{J.-H. Xue}, \bibinfo{author}{S.~Ruan},
  \bibinfo{author}{B.~Moretti}, \bibinfo{author}{M.~Revenu},
  \bibinfo{author}{D.~Bloyet}, \bibinfo{title}{Knowledge-based segmentation and
  labeling of brain structures from MRI images}, \bibinfo{journal}{Pattern
  recognition letters} \bibinfo{volume}{22}~(\bibinfo{number}{3-4})
  (\bibinfo{year}{2001}) \bibinfo{pages}{395--405}.

\bibitem[{Gui et~al.(2012)Gui, Lisowski, Faundez, H{\"u}ppi, Lazeyras, and
  Kocher}]{gui2012morphology}
\bibinfo{author}{L.~Gui}, \bibinfo{author}{R.~Lisowski},
  \bibinfo{author}{T.~Faundez}, \bibinfo{author}{P.~S. H{\"u}ppi},
  \bibinfo{author}{F.~Lazeyras}, \bibinfo{author}{M.~Kocher},
  \bibinfo{title}{Morphology-driven automatic segmentation of MR images of the
  neonatal brain}, \bibinfo{journal}{Medical Image Analysis}
  \bibinfo{volume}{16}~(\bibinfo{number}{8}) (\bibinfo{year}{2012})
  \bibinfo{pages}{1565--1579}.

\bibitem[{Huang et~al.(2009)Huang, Abugharbieh, Tam, Initiative
  et~al.}]{huang2009hybrid}
\bibinfo{author}{A.~Huang}, \bibinfo{author}{R.~Abugharbieh},
  \bibinfo{author}{R.~Tam}, \bibinfo{author}{A.~D.~N. Initiative}, et~al.,
  \bibinfo{title}{A hybrid geometric--statistical deformable model for
  automated 3-D segmentation in brain MRI}, \bibinfo{journal}{IEEE Transactions
  on Biomedical Engineering} \bibinfo{volume}{56}~(\bibinfo{number}{7})
  (\bibinfo{year}{2009}) \bibinfo{pages}{1838--1848}.

\bibitem[{Kapur et~al.(1996)Kapur, Grimson, Wells~III, and
  Kikinis}]{kapur1996segmentation}
\bibinfo{author}{T.~Kapur}, \bibinfo{author}{W.~E.~L. Grimson},
  \bibinfo{author}{W.~M. Wells~III}, \bibinfo{author}{R.~Kikinis},
  \bibinfo{title}{Segmentation of brain tissue from magnetic resonance images},
  \bibinfo{journal}{Medical Image Analysis}
  \bibinfo{volume}{1}~(\bibinfo{number}{2}) (\bibinfo{year}{1996})
  \bibinfo{pages}{109--127}.

\bibitem[{Tang et~al.(2000)Tang, Wu, Ma, Gallagher, Perera, and
  Zhuang}]{tang2000mri}
\bibinfo{author}{H.~Tang}, \bibinfo{author}{E.~Wu}, \bibinfo{author}{Q.~Ma},
  \bibinfo{author}{D.~Gallagher}, \bibinfo{author}{G.~Perera},
  \bibinfo{author}{T.~Zhuang}, \bibinfo{title}{MRI brain image segmentation by
  multi-resolution edge detection and region selection},
  \bibinfo{journal}{Computerized Medical Imaging and Graphics}
  \bibinfo{volume}{24}~(\bibinfo{number}{6}) (\bibinfo{year}{2000})
  \bibinfo{pages}{349--357}.

\bibitem[{Wang et~al.(2010)Wang, Chen, Pan, Hong, and Xia}]{wang2010level}
\bibinfo{author}{L.~Wang}, \bibinfo{author}{Y.~Chen}, \bibinfo{author}{X.~Pan},
  \bibinfo{author}{X.~Hong}, \bibinfo{author}{D.~Xia}, \bibinfo{title}{Level
  set segmentation of brain magnetic resonance images based on local Gaussian
  distribution fitting energy}, \bibinfo{journal}{Journal of Neuroscience
  Methods} \bibinfo{volume}{188}~(\bibinfo{number}{2}) (\bibinfo{year}{2010})
  \bibinfo{pages}{316--325}.

\bibitem[{Chen et~al.(2008)Chen, Qiu, and Ruan}]{chen2008fuzzy}
\bibinfo{author}{Z.~Chen}, \bibinfo{author}{T.~Qiu}, \bibinfo{author}{S.~Ruan},
  \bibinfo{title}{Fuzzy adaptive level set algorithm for brain tissue
  segmentation}, in: \bibinfo{booktitle}{2008 9th International Conference on
  Signal Processing}, \bibinfo{organization}{IEEE},
  \bibinfo{pages}{1047--1050}, \bibinfo{year}{2008}.

\bibitem[{Wang et~al.(2013)Wang, Shi, Yap, Lin, Gilmore, and
  Shen}]{wang2013longitudinally}
\bibinfo{author}{L.~Wang}, \bibinfo{author}{F.~Shi}, \bibinfo{author}{P.-T.
  Yap}, \bibinfo{author}{W.~Lin}, \bibinfo{author}{J.~H. Gilmore},
  \bibinfo{author}{D.~Shen}, \bibinfo{title}{Longitudinally guided level sets
  for consistent tissue segmentation of neonates}, \bibinfo{journal}{Human
  brain mapping} \bibinfo{volume}{34}~(\bibinfo{number}{4})
  (\bibinfo{year}{2013}) \bibinfo{pages}{956--972}.

\bibitem[{Song et~al.(2006)Song, Tustison, Avants, and
  Gee}]{song2006integrated}
\bibinfo{author}{Z.~Song}, \bibinfo{author}{N.~Tustison},
  \bibinfo{author}{B.~Avants}, \bibinfo{author}{J.~C. Gee},
  \bibinfo{title}{Integrated graph cuts for brain MRI segmentation}, in:
  \bibinfo{booktitle}{International Conference on Medical Image Computing and
  Computer-Assisted Intervention}, \bibinfo{pages}{831--838},
  \bibinfo{year}{2006}.

\bibitem[{Kalavathi(2013)}]{kalavathi2013brain}
\bibinfo{author}{P.~Kalavathi}, \bibinfo{title}{Brain tissue segmentation in MR
  brain images using multiple Otsu's thresholding technique}, in:
  \bibinfo{booktitle}{Proceedings of the International Conference on Computer
  Science and Education}, \bibinfo{organization}{IEEE},
  \bibinfo{pages}{639--642}, \bibinfo{year}{2013}.

\bibitem[{De~Boer et~al.(2009)De~Boer, Vrooman, Van Der~Lijn, Vernooij, Ikram,
  Van Der~Lugt, Breteler, and Niessen}]{de2009white}
\bibinfo{author}{R.~De~Boer}, \bibinfo{author}{H.~A. Vrooman},
  \bibinfo{author}{F.~Van Der~Lijn}, \bibinfo{author}{M.~W. Vernooij},
  \bibinfo{author}{M.~A. Ikram}, \bibinfo{author}{A.~Van Der~Lugt},
  \bibinfo{author}{M.~M. Breteler}, \bibinfo{author}{W.~J. Niessen},
  \bibinfo{title}{White matter lesion extension to automatic brain tissue
  segmentation on MRI}, \bibinfo{journal}{NeuroImage}
  \bibinfo{volume}{45}~(\bibinfo{number}{4}) (\bibinfo{year}{2009})
  \bibinfo{pages}{1151--1161}.

\bibitem[{Sathya and Kayalvizhi(2011)}]{sathya2011optimal}
\bibinfo{author}{P.~D. Sathya}, \bibinfo{author}{R.~Kayalvizhi},
  \bibinfo{title}{Optimal segmentation of brain MRI based on adaptive bacterial
  foraging algorithm}, \bibinfo{journal}{Neurocomputing}
  \bibinfo{volume}{74}~(\bibinfo{number}{14-15}) (\bibinfo{year}{2011})
  \bibinfo{pages}{2299--2313}.

\bibitem[{Kapur et~al.(1985)Kapur, Sahoo, and Wong}]{kapur1985new}
\bibinfo{author}{J.~N. Kapur}, \bibinfo{author}{P.~K. Sahoo},
  \bibinfo{author}{A.~K. Wong}, \bibinfo{title}{A new method for gray-level
  picture thresholding using the entropy of the histogram},
  \bibinfo{journal}{Computer vision, graphics, and image processing}
  \bibinfo{volume}{29}~(\bibinfo{number}{3}) (\bibinfo{year}{1985})
  \bibinfo{pages}{273--285}.

\bibitem[{Constantin et~al.(2010)Constantin, Bajcsy, and
  Nelson}]{constantin2010unsupervised}
\bibinfo{author}{A.~A. Constantin}, \bibinfo{author}{B.~R. Bajcsy},
  \bibinfo{author}{C.~S. Nelson}, \bibinfo{title}{Unsupervised segmentation of
  brain tissue in multivariate MRI}, in: \bibinfo{booktitle}{2010 IEEE
  international symposium on biomedical imaging: from nano to macro},
  \bibinfo{organization}{IEEE}, \bibinfo{pages}{89--92}, \bibinfo{year}{2010}.

\bibitem[{Barra and Boire(2000)}]{barra2000tissue}
\bibinfo{author}{V.~Barra}, \bibinfo{author}{J.-Y. Boire},
  \bibinfo{title}{Tissue segmentation on MR images of the brain by
  possibilistic clustering on a 3D wavelet representation},
  \bibinfo{journal}{Journal of Magnetic Resonance Imaging: An Official Journal
  of the International Society for Magnetic Resonance in Medicine}
  \bibinfo{volume}{11}~(\bibinfo{number}{3}) (\bibinfo{year}{2000})
  \bibinfo{pages}{267--278}.

\bibitem[{Shen et~al.(2005)Shen, Sandham, Granat, and Sterr}]{shen2005mri}
\bibinfo{author}{S.~Shen}, \bibinfo{author}{W.~Sandham},
  \bibinfo{author}{M.~Granat}, \bibinfo{author}{A.~Sterr}, \bibinfo{title}{MRI
  fuzzy segmentation of brain tissue using neighborhood attraction with
  neural-network optimization}, \bibinfo{journal}{IEEE Transactions on
  Information Technology in Biomedicine}
  \bibinfo{volume}{9}~(\bibinfo{number}{3}) (\bibinfo{year}{2005})
  \bibinfo{pages}{459--467}.

\bibitem[{Halder and Talukdar(2019)}]{halder2019brain}
\bibinfo{author}{A.~Halder}, \bibinfo{author}{N.~A. Talukdar},
  \bibinfo{title}{Brain tissue segmentation using improved kernelized
  rough-fuzzy C-means with spatio-contextual information from MRI},
  \bibinfo{journal}{Magnetic Resonance Imaging} \bibinfo{volume}{62}
  (\bibinfo{year}{2019}) \bibinfo{pages}{129--151}.

\bibitem[{Rajapakse et~al.(1996)Rajapakse, Giedd, DeCarli, Snell, McLaughlin,
  Vauss, Krain, Hamburger, and Rapoport}]{rajapakse1996technique}
\bibinfo{author}{J.~C. Rajapakse}, \bibinfo{author}{J.~N. Giedd},
  \bibinfo{author}{C.~DeCarli}, \bibinfo{author}{J.~W. Snell},
  \bibinfo{author}{A.~McLaughlin}, \bibinfo{author}{Y.~C. Vauss},
  \bibinfo{author}{A.~L. Krain}, \bibinfo{author}{S.~Hamburger},
  \bibinfo{author}{J.~L. Rapoport}, \bibinfo{title}{A technique for
  single-channel MR brain tissue segmentation: application to a pediatric
  sample}, \bibinfo{journal}{Magnetic Resonance Imaging}
  \bibinfo{volume}{14}~(\bibinfo{number}{9}) (\bibinfo{year}{1996})
  \bibinfo{pages}{1053--1065}.

\bibitem[{Silva(2007)}]{da2007dirichlet}
\bibinfo{author}{A.~R. F.~d. Silva}, \bibinfo{title}{A Dirichlet process
  mixture model for brain MRI tissue classification}, \bibinfo{journal}{Medical
  Image Analysis} \bibinfo{volume}{11}~(\bibinfo{number}{2})
  (\bibinfo{year}{2007}) \bibinfo{pages}{169--182}.

\bibitem[{Kouw et~al.(2019)Kouw, {\O}rting, Petersen, Pedersen, and
  de~Bruijne}]{kouw2019cross}
\bibinfo{author}{W.~M. Kouw}, \bibinfo{author}{S.~N. {\O}rting},
  \bibinfo{author}{J.~Petersen}, \bibinfo{author}{K.~S. Pedersen},
  \bibinfo{author}{M.~de~Bruijne}, \bibinfo{title}{A cross-center smoothness
  prior for variational Bayesian brain tissue segmentation}, in:
  \bibinfo{booktitle}{International Conference on Information Processing in
  Medical Imaging}, \bibinfo{pages}{360--371}, \bibinfo{year}{2019}.

\bibitem[{Yousefi et~al.(2012)Yousefi, Azmi, and Zahedi}]{yousefi2012brain}
\bibinfo{author}{S.~Yousefi}, \bibinfo{author}{R.~Azmi},
  \bibinfo{author}{M.~Zahedi}, \bibinfo{title}{Brain tissue segmentation in MR
  images based on a hybrid of MRF and social algorithms},
  \bibinfo{journal}{Medical Image Analysis}
  \bibinfo{volume}{16}~(\bibinfo{number}{4}) (\bibinfo{year}{2012})
  \bibinfo{pages}{840--848}.

\bibitem[{Wells et~al.(1996)Wells, Grimson, Kikinis, and
  Jolesz}]{wells1996adaptive}
\bibinfo{author}{W.~M. Wells}, \bibinfo{author}{W.~E.~L. Grimson},
  \bibinfo{author}{R.~Kikinis}, \bibinfo{author}{F.~A. Jolesz},
  \bibinfo{title}{Adaptive segmentation of MRI data}, \bibinfo{journal}{IEEE
  transactions on medical imaging} \bibinfo{volume}{15}~(\bibinfo{number}{4})
  (\bibinfo{year}{1996}) \bibinfo{pages}{429--442}.

\bibitem[{Van~Leemput et~al.(1999)Van~Leemput, Maes, Vandermeulen, and
  Suetens}]{van1999automated}
\bibinfo{author}{K.~Van~Leemput}, \bibinfo{author}{F.~Maes},
  \bibinfo{author}{D.~Vandermeulen}, \bibinfo{author}{P.~Suetens},
  \bibinfo{title}{Automated model-based tissue classification of MR images of
  the brain}, \bibinfo{journal}{IEEE transactions on medical imaging}
  \bibinfo{volume}{18}~(\bibinfo{number}{10}) (\bibinfo{year}{1999})
  \bibinfo{pages}{897--908}.

\bibitem[{Ashburner and Friston(2005)}]{ashburner2005unified}
\bibinfo{author}{J.~Ashburner}, \bibinfo{author}{K.~J. Friston},
  \bibinfo{title}{Unified segmentation}, \bibinfo{journal}{Neuroimage}
  \bibinfo{volume}{26}~(\bibinfo{number}{3}) (\bibinfo{year}{2005})
  \bibinfo{pages}{839--851}.

\bibitem[{Liang et~al.(1994)Liang, MacFall, and
  Harrington}]{liang1994parameter}
\bibinfo{author}{Z.~Liang}, \bibinfo{author}{J.~R. MacFall},
  \bibinfo{author}{D.~P. Harrington}, \bibinfo{title}{Parameter estimation and
  tissue segmentation from multispectral MR images}, \bibinfo{journal}{IEEE
  transactions on Medical Imaging} \bibinfo{volume}{13}~(\bibinfo{number}{3})
  (\bibinfo{year}{1994}) \bibinfo{pages}{441--449}.

\bibitem[{Kong et~al.(2014)Kong, Deng, and Dai}]{kong2014discriminative}
\bibinfo{author}{Y.~Kong}, \bibinfo{author}{Y.~Deng}, \bibinfo{author}{Q.~Dai},
  \bibinfo{title}{Discriminative clustering and feature selection for brain MRI
  segmentation}, \bibinfo{journal}{IEEE Signal Processing Letters}
  \bibinfo{volume}{22}~(\bibinfo{number}{5}) (\bibinfo{year}{2014})
  \bibinfo{pages}{573--577}.

\bibitem[{Paul et~al.(2015)Paul, Varghese, Purushothaman, and
  Singh}]{paul2015automated}
\bibinfo{author}{G.~Paul}, \bibinfo{author}{T.~Varghese},
  \bibinfo{author}{K.~Purushothaman}, \bibinfo{author}{A.~Singh},
  \bibinfo{title}{Automated Segmentation of MR Images by Implementing Multi SVM
  Technique}, in: \bibinfo{booktitle}{Power Electronics and Renewable Energy
  Systems}, \bibinfo{publisher}{Springer}, \bibinfo{pages}{1509--1516},
  \bibinfo{year}{2015}.

\bibitem[{Vrooman et~al.(2007)Vrooman, Cocosco, van~der Lijn, Stokking, Ikram,
  Vernooij, Breteler, and Niessen}]{vrooman2007multi}
\bibinfo{author}{H.~A. Vrooman}, \bibinfo{author}{C.~A. Cocosco},
  \bibinfo{author}{F.~van~der Lijn}, \bibinfo{author}{R.~Stokking},
  \bibinfo{author}{M.~A. Ikram}, \bibinfo{author}{M.~W. Vernooij},
  \bibinfo{author}{M.~M. Breteler}, \bibinfo{author}{W.~J. Niessen},
  \bibinfo{title}{Multi-spectral brain tissue segmentation using automatically
  trained k-Nearest-Neighbor classification}, \bibinfo{journal}{NeuroImage}
  \bibinfo{volume}{37}~(\bibinfo{number}{1}) (\bibinfo{year}{2007})
  \bibinfo{pages}{71--81}.

\bibitem[{Anbeek et~al.(2005)Anbeek, Vincken, Van~Bochove, Van~Osch, and
  van~der Grond}]{anbeek2005probabilistic}
\bibinfo{author}{P.~Anbeek}, \bibinfo{author}{K.~L. Vincken},
  \bibinfo{author}{G.~S. Van~Bochove}, \bibinfo{author}{M.~J. Van~Osch},
  \bibinfo{author}{J.~van~der Grond}, \bibinfo{title}{Probabilistic
  segmentation of brain tissue in MR imaging}, \bibinfo{journal}{Neuroimage}
  \bibinfo{volume}{27}~(\bibinfo{number}{4}) (\bibinfo{year}{2005})
  \bibinfo{pages}{795--804}.

\bibitem[{Anbeek et~al.(2008)Anbeek, Vincken, Groenendaal, Koeman, Van~Osch,
  and Van~der Grond}]{anbeek2008probabilistic}
\bibinfo{author}{P.~Anbeek}, \bibinfo{author}{K.~L. Vincken},
  \bibinfo{author}{F.~Groenendaal}, \bibinfo{author}{A.~Koeman},
  \bibinfo{author}{M.~J. Van~Osch}, \bibinfo{author}{J.~Van~der Grond},
  \bibinfo{title}{Probabilistic brain tissue segmentation in neonatal magnetic
  resonance imaging}, \bibinfo{journal}{Pediatric research}
  \bibinfo{volume}{63}~(\bibinfo{number}{2}) (\bibinfo{year}{2008})
  \bibinfo{pages}{158--163}.

\bibitem[{van Opbroek et~al.(2013)van Opbroek, van~der Lijn, and
  de~Bruijne}]{van2013automated}
\bibinfo{author}{A.~van Opbroek}, \bibinfo{author}{F.~van~der Lijn},
  \bibinfo{author}{M.~de~Bruijne}, \bibinfo{title}{Automated brain-tissue
  segmentation by multi-feature SVM classification}, in:
  \bibinfo{booktitle}{Proceedings of the MICCAI Workshops—The MICCAI Grand
  Challenge on MR Brain Image Segmentation (MRBrainS’13)},
  \bibinfo{year}{2013}.

\bibitem[{Yi et~al.(2009)Yi, Criminisi, Shotton, and
  Blake}]{yi2009discriminative}
\bibinfo{author}{Z.~Yi}, \bibinfo{author}{A.~Criminisi},
  \bibinfo{author}{J.~Shotton}, \bibinfo{author}{A.~Blake},
  \bibinfo{title}{Discriminative, semantic segmentation of brain tissue in MR
  images}, in: \bibinfo{booktitle}{International Conference on Medical Image
  Computing and Computer-Assisted Intervention},
  \bibinfo{organization}{Springer}, \bibinfo{pages}{558--565},
  \bibinfo{year}{2009}.

\bibitem[{Liu et~al.(2014)Liu, Li, Wang, Wu, Liu, and Pan}]{liu2014survey}
\bibinfo{author}{J.~Liu}, \bibinfo{author}{M.~Li}, \bibinfo{author}{J.~Wang},
  \bibinfo{author}{F.~Wu}, \bibinfo{author}{T.~Liu}, \bibinfo{author}{Y.~Pan},
  \bibinfo{title}{A survey of MRI-based brain tumor segmentation methods},
  \bibinfo{journal}{Tsinghua Science and Technology}
  \bibinfo{volume}{19}~(\bibinfo{number}{6}) (\bibinfo{year}{2014})
  \bibinfo{pages}{578--595}.

\bibitem[{Chudasama and Robbins(2006)}]{chudasama2006functions}
\bibinfo{author}{Y.~Chudasama}, \bibinfo{author}{T.~Robbins},
  \bibinfo{title}{Functions of frontostriatal systems in cognition: comparative
  neuropsychopharmacological studies in rats, monkeys and humans},
  \bibinfo{journal}{Biological Psychology}
  \bibinfo{volume}{73}~(\bibinfo{number}{1}) (\bibinfo{year}{2006})
  \bibinfo{pages}{19--38}.

\bibitem[{Apostolova et~al.(2010)Apostolova, Mosconi, Thompson, Green, Hwang,
  Ramirez, Mistur, Tsui, and de~Leon}]{apostolova2010subregional}
\bibinfo{author}{L.~G. Apostolova}, \bibinfo{author}{L.~Mosconi},
  \bibinfo{author}{P.~M. Thompson}, \bibinfo{author}{A.~E. Green},
  \bibinfo{author}{K.~S. Hwang}, \bibinfo{author}{A.~Ramirez},
  \bibinfo{author}{R.~Mistur}, \bibinfo{author}{W.~H. Tsui},
  \bibinfo{author}{M.~J. de~Leon}, \bibinfo{title}{Subregional hippocampal
  atrophy predicts Alzheimer's dementia in the cognitively normal},
  \bibinfo{journal}{Neurobiology of Aging}
  \bibinfo{volume}{31}~(\bibinfo{number}{7}) (\bibinfo{year}{2010})
  \bibinfo{pages}{1077--1088}.

\bibitem[{Sakalauskas et~al.(2010)Sakalauskas, Luko{\v{s}}evi{\v{c}}ius, and
  Lau{\v{c}}kait{\.e}}]{sakalauskas2010transcranial}
\bibinfo{author}{A.~Sakalauskas},
  \bibinfo{author}{A.~Luko{\v{s}}evi{\v{c}}ius},
  \bibinfo{author}{K.~Lau{\v{c}}kait{\.e}}, \bibinfo{title}{Transcranial
  echoscopy for diagnostic of Parkinson disease: technical constraints and
  possibilities}, \bibinfo{journal}{Ultragarsas}
  \bibinfo{volume}{65}~(\bibinfo{number}{4}) (\bibinfo{year}{2010})
  \bibinfo{pages}{47--50}.

\bibitem[{Kushibar et~al.(2018)Kushibar, Valverde, Gonz{\'a}lez-Vill{\`a},
  Bernal, Cabezas, Oliver, and Llad{\'o}}]{kushibar2018automated}
\bibinfo{author}{K.~Kushibar}, \bibinfo{author}{S.~Valverde},
  \bibinfo{author}{S.~Gonz{\'a}lez-Vill{\`a}}, \bibinfo{author}{J.~Bernal},
  \bibinfo{author}{M.~Cabezas}, \bibinfo{author}{A.~Oliver},
  \bibinfo{author}{X.~Llad{\'o}}, \bibinfo{title}{Automated sub-cortical brain
  structure segmentation combining spatial and deep convolutional features},
  \bibinfo{journal}{Medical Image Analysis} \bibinfo{volume}{48}
  (\bibinfo{year}{2018}) \bibinfo{pages}{177--186}.

\bibitem[{Xia et~al.(2007)Xia, Bettinger, Shen, and Reiss}]{xia2007automatic}
\bibinfo{author}{Y.~Xia}, \bibinfo{author}{K.~Bettinger},
  \bibinfo{author}{L.~Shen}, \bibinfo{author}{A.~L. Reiss},
  \bibinfo{title}{Automatic segmentation of the caudate nucleus from human
  brain MR images}, \bibinfo{journal}{IEEE Transactions on Medical Imaging}
  \bibinfo{volume}{26}~(\bibinfo{number}{4}) (\bibinfo{year}{2007})
  \bibinfo{pages}{509--517}.

\bibitem[{Pipitone et~al.(2014)Pipitone, Park, Winterburn, Lett, Lerch,
  Pruessner, Lepage, Voineskos, Chakravarty, Initiative
  et~al.}]{pipitone2014multi}
\bibinfo{author}{J.~Pipitone}, \bibinfo{author}{M.~T.~M. Park},
  \bibinfo{author}{J.~Winterburn}, \bibinfo{author}{T.~A. Lett},
  \bibinfo{author}{J.~P. Lerch}, \bibinfo{author}{J.~C. Pruessner},
  \bibinfo{author}{M.~Lepage}, \bibinfo{author}{A.~N. Voineskos},
  \bibinfo{author}{M.~M. Chakravarty}, \bibinfo{author}{A.~D.~N. Initiative},
  et~al., \bibinfo{title}{Multi-atlas segmentation of the whole hippocampus and
  subfields using multiple automatically generated templates},
  \bibinfo{journal}{NeuroImage} \bibinfo{volume}{101} (\bibinfo{year}{2014})
  \bibinfo{pages}{494--512}.

\bibitem[{Fischl et~al.(2002)Fischl, Salat, Busa, Albert, Dieterich,
  Haselgrove, Van Der~Kouwe, Killiany, Kennedy, Klaveness
  et~al.}]{fischl2002whole}
\bibinfo{author}{B.~Fischl}, \bibinfo{author}{D.~H. Salat},
  \bibinfo{author}{E.~Busa}, \bibinfo{author}{M.~Albert},
  \bibinfo{author}{M.~Dieterich}, \bibinfo{author}{C.~Haselgrove},
  \bibinfo{author}{A.~Van Der~Kouwe}, \bibinfo{author}{R.~Killiany},
  \bibinfo{author}{D.~Kennedy}, \bibinfo{author}{S.~Klaveness}, et~al.,
  \bibinfo{title}{Whole brain segmentation: automated labeling of
  neuroanatomical structures in the human brain}, \bibinfo{journal}{Neuron}
  \bibinfo{volume}{33}~(\bibinfo{number}{3}) (\bibinfo{year}{2002})
  \bibinfo{pages}{341--355}.

\bibitem[{Patenaude et~al.(2011)Patenaude, Smith, Kennedy, and
  Jenkinson}]{patenaude2011bayesian}
\bibinfo{author}{B.~Patenaude}, \bibinfo{author}{S.~M. Smith},
  \bibinfo{author}{D.~N. Kennedy}, \bibinfo{author}{M.~Jenkinson},
  \bibinfo{title}{A Bayesian model of shape and appearance for subcortical
  brain segmentation}, \bibinfo{journal}{NeuroImage}
  \bibinfo{volume}{56}~(\bibinfo{number}{3}) (\bibinfo{year}{2011})
  \bibinfo{pages}{907--922}.

\bibitem[{Landman and Warfield(2012)}]{landman2012miccai}
\bibinfo{author}{B.~Landman}, \bibinfo{author}{S.~Warfield},
  \bibinfo{title}{MICCAI 2012 workshop on multi-atlas labeling}, in:
  \bibinfo{booktitle}{International Conference on Medical Image Computing and
  Computer-Assisted Intervention}, \bibinfo{year}{2012}.

\bibitem[{Sabuncu et~al.(2010)Sabuncu, Yeo, Van~Leemput, Fischl, and
  Golland}]{sabuncu2010generative}
\bibinfo{author}{M.~R. Sabuncu}, \bibinfo{author}{B.~T. Yeo},
  \bibinfo{author}{K.~Van~Leemput}, \bibinfo{author}{B.~Fischl},
  \bibinfo{author}{P.~Golland}, \bibinfo{title}{A generative model for image
  segmentation based on label fusion}, \bibinfo{journal}{IEEE Transactions on
  Medical Imaging} \bibinfo{volume}{29}~(\bibinfo{number}{10})
  (\bibinfo{year}{2010}) \bibinfo{pages}{1714--1729}.

\bibitem[{Commowick et~al.(2012)Commowick, Akhondi-Asl, and
  Warfield}]{commowick2012estimating}
\bibinfo{author}{O.~Commowick}, \bibinfo{author}{A.~Akhondi-Asl},
  \bibinfo{author}{S.~K. Warfield}, \bibinfo{title}{Estimating a reference
  standard segmentation with spatially varying performance parameters: Local
  MAP STAPLE}, \bibinfo{journal}{IEEE Transactions on Medical Imaging}
  \bibinfo{volume}{31}~(\bibinfo{number}{8}) (\bibinfo{year}{2012})
  \bibinfo{pages}{1593--1606}.

\bibitem[{Huo et~al.(2016)Huo, Plassard, Carass, Resnick, Pham, Prince, and
  Landman}]{huo2016consistent}
\bibinfo{author}{Y.~Huo}, \bibinfo{author}{A.~J. Plassard},
  \bibinfo{author}{A.~Carass}, \bibinfo{author}{S.~M. Resnick},
  \bibinfo{author}{D.~L. Pham}, \bibinfo{author}{J.~L. Prince},
  \bibinfo{author}{B.~A. Landman}, \bibinfo{title}{Consistent cortical
  reconstruction and multi-atlas brain segmentation},
  \bibinfo{journal}{NeuroImage} \bibinfo{volume}{138} (\bibinfo{year}{2016})
  \bibinfo{pages}{197--210}.

\bibitem[{Puonti et~al.(2016)Puonti, Iglesias, and
  Van~Leemput}]{puonti2016fast}
\bibinfo{author}{O.~Puonti}, \bibinfo{author}{J.~E. Iglesias},
  \bibinfo{author}{K.~Van~Leemput}, \bibinfo{title}{Fast and sequence-adaptive
  whole-brain segmentation using parametric Bayesian modeling},
  \bibinfo{journal}{NeuroImage} \bibinfo{volume}{143} (\bibinfo{year}{2016})
  \bibinfo{pages}{235--249}.

\bibitem[{Asman and Landman(2014)}]{asman2014hierarchical}
\bibinfo{author}{A.~J. Asman}, \bibinfo{author}{B.~A. Landman},
  \bibinfo{title}{Hierarchical performance estimation in the statistical label
  fusion framework}, \bibinfo{journal}{Medical Image Analysis}
  \bibinfo{volume}{18}~(\bibinfo{number}{7}) (\bibinfo{year}{2014})
  \bibinfo{pages}{1070--1081}.

\bibitem[{Chen et~al.(2018{\natexlab{b}})Chen, Dou, Yu, Qin, and
  Heng}]{chen2018voxresnet}
\bibinfo{author}{H.~Chen}, \bibinfo{author}{Q.~Dou}, \bibinfo{author}{L.~Yu},
  \bibinfo{author}{J.~Qin}, \bibinfo{author}{P.-A. Heng},
  \bibinfo{title}{VoxResNet: Deep voxelwise residual networks for brain
  segmentation from 3D MR images}, \bibinfo{journal}{NeuroImage}
  \bibinfo{volume}{170} (\bibinfo{year}{2018}{\natexlab{b}})
  \bibinfo{pages}{446--455}.

\bibitem[{Dolz et~al.(2018{\natexlab{a}})Dolz, Gopinath, Yuan, Lombaert,
  Desrosiers, and Ayed}]{dolz2018hyperdense}
\bibinfo{author}{J.~Dolz}, \bibinfo{author}{K.~Gopinath},
  \bibinfo{author}{J.~Yuan}, \bibinfo{author}{H.~Lombaert},
  \bibinfo{author}{C.~Desrosiers}, \bibinfo{author}{I.~B. Ayed},
  \bibinfo{title}{HyperDense-Net: A hyper-densely connected CNN for multi-modal
  image segmentation}, \bibinfo{journal}{IEEE Transactions on Medical Imaging}
  \bibinfo{volume}{38}~(\bibinfo{number}{5})
  (\bibinfo{year}{2018}{\natexlab{a}}) \bibinfo{pages}{1116--1126}.

\bibitem[{Wachinger et~al.(2018)Wachinger, Reuter, and
  Klein}]{wachinger2018deepnat}
\bibinfo{author}{C.~Wachinger}, \bibinfo{author}{M.~Reuter},
  \bibinfo{author}{T.~Klein}, \bibinfo{title}{DeepNAT: Deep convolutional
  neural network for segmenting neuroanatomy}, \bibinfo{journal}{NeuroImage}
  \bibinfo{volume}{170} (\bibinfo{year}{2018}) \bibinfo{pages}{434--445}.

\bibitem[{Billot et~al.(2020)Billot, Greve, Leemput, Fischl, Iglesias, and
  Dalca}]{billot2020a}
\bibinfo{author}{B.~Billot}, \bibinfo{author}{D.~N. Greve},
  \bibinfo{author}{K.~V. Leemput}, \bibinfo{author}{B.~Fischl},
  \bibinfo{author}{J.~E. Iglesias}, \bibinfo{author}{A.~V. Dalca},
  \bibinfo{title}{A learning strategy for contrast-agnostic {\{}MRI{\}}
  segmentation}, in: \bibinfo{booktitle}{International Conference on Medical
  Imaging with Deep Learning}, \bibinfo{year}{2020}.

\bibitem[{Girshick et~al.(2014)Girshick, Donahue, Darrell, and
  Malik}]{girshick2014rich}
\bibinfo{author}{R.~Girshick}, \bibinfo{author}{J.~Donahue},
  \bibinfo{author}{T.~Darrell}, \bibinfo{author}{J.~Malik},
  \bibinfo{title}{Rich feature hierarchies for accurate object detection and
  semantic segmentation}, in: \bibinfo{booktitle}{Proceedings of the IEEE
  Conference on Computer Vision and Pattern Recognition},
  \bibinfo{pages}{580--587}, \bibinfo{year}{2014}.

\bibitem[{Chen et~al.(2017)Chen, Papandreou, Kokkinos, Murphy, and
  Yuille}]{chen2017deeplab}
\bibinfo{author}{L.-C. Chen}, \bibinfo{author}{G.~Papandreou},
  \bibinfo{author}{I.~Kokkinos}, \bibinfo{author}{K.~Murphy},
  \bibinfo{author}{A.~L. Yuille}, \bibinfo{title}{Deeplab: Semantic image
  segmentation with deep convolutional nets, atrous convolution, and fully
  connected crfs}, \bibinfo{journal}{IEEE transactions on pattern analysis and
  machine intelligence} \bibinfo{volume}{40}~(\bibinfo{number}{4})
  (\bibinfo{year}{2017}) \bibinfo{pages}{834--848}.

\bibitem[{Chen et~al.(2016)Chen, Yang, Wang, Xu, and
  Yuille}]{chen2016attention}
\bibinfo{author}{L.-C. Chen}, \bibinfo{author}{Y.~Yang},
  \bibinfo{author}{J.~Wang}, \bibinfo{author}{W.~Xu}, \bibinfo{author}{A.~L.
  Yuille}, \bibinfo{title}{Attention to scale: Scale-aware semantic image
  segmentation}, in: \bibinfo{booktitle}{Proceedings of the IEEE conference on
  computer vision and pattern recognition}, \bibinfo{pages}{3640--3649},
  \bibinfo{year}{2016}.

\bibitem[{Zhao et~al.(2017)Zhao, Shi, Qi, Wang, and Jia}]{zhao2017pyramid}
\bibinfo{author}{H.~Zhao}, \bibinfo{author}{J.~Shi}, \bibinfo{author}{X.~Qi},
  \bibinfo{author}{X.~Wang}, \bibinfo{author}{J.~Jia}, \bibinfo{title}{Pyramid
  scene parsing network}, in: \bibinfo{booktitle}{Proceedings of the IEEE
  conference on computer vision and pattern recognition},
  \bibinfo{pages}{2881--2890}, \bibinfo{year}{2017}.

\bibitem[{Milletari et~al.(2016)Milletari, Navab, and Ahmadi}]{milletari2016v}
\bibinfo{author}{F.~Milletari}, \bibinfo{author}{N.~Navab},
  \bibinfo{author}{S.-A. Ahmadi}, \bibinfo{title}{V-net: Fully convolutional
  neural networks for volumetric medical image segmentation}, in:
  \bibinfo{booktitle}{International Conference on 3D Vision},
  \bibinfo{pages}{565--571}, \bibinfo{year}{2016}.

\bibitem[{Badrinarayanan et~al.(2017)Badrinarayanan, Kendall, and
  Cipolla}]{badrinarayanan2017segnet}
\bibinfo{author}{V.~Badrinarayanan}, \bibinfo{author}{A.~Kendall},
  \bibinfo{author}{R.~Cipolla}, \bibinfo{title}{Segnet: A deep convolutional
  encoder-decoder architecture for image segmentation}, \bibinfo{journal}{IEEE
  transactions on pattern analysis and machine intelligence}
  \bibinfo{volume}{39}~(\bibinfo{number}{12}) (\bibinfo{year}{2017})
  \bibinfo{pages}{2481--2495}.

\bibitem[{Noh et~al.(2015)Noh, Hong, and Han}]{noh2015learning}
\bibinfo{author}{H.~Noh}, \bibinfo{author}{S.~Hong}, \bibinfo{author}{B.~Han},
  \bibinfo{title}{Learning deconvolution network for semantic segmentation},
  in: \bibinfo{booktitle}{Proceedings of the IEEE international conference on
  computer vision}, \bibinfo{pages}{1520--1528}, \bibinfo{year}{2015}.

\bibitem[{Lin et~al.(2017)Lin, Milan, Shen, and Reid}]{lin2017refinenet}
\bibinfo{author}{G.~Lin}, \bibinfo{author}{A.~Milan},
  \bibinfo{author}{C.~Shen}, \bibinfo{author}{I.~Reid},
  \bibinfo{title}{Refinenet: Multi-path refinement networks for high-resolution
  semantic segmentation}, in: \bibinfo{booktitle}{Proceedings of the IEEE
  conference on computer vision and pattern recognition},
  \bibinfo{pages}{1925--1934}, \bibinfo{year}{2017}.

\bibitem[{Fu et~al.(2019)Fu, Liu, Wang, Zhou, Wang, and Lu}]{fu2019stacked}
\bibinfo{author}{J.~Fu}, \bibinfo{author}{J.~Liu}, \bibinfo{author}{Y.~Wang},
  \bibinfo{author}{J.~Zhou}, \bibinfo{author}{C.~Wang},
  \bibinfo{author}{H.~Lu}, \bibinfo{title}{Stacked deconvolutional network for
  semantic segmentation}, \bibinfo{journal}{IEEE Transactions on Image
  Processing} .

\bibitem[{Chen et~al.(2018{\natexlab{c}})Chen, Zhu, Papandreou, Schroff, and
  Adam}]{chen2018encoder}
\bibinfo{author}{L.-C. Chen}, \bibinfo{author}{Y.~Zhu},
  \bibinfo{author}{G.~Papandreou}, \bibinfo{author}{F.~Schroff},
  \bibinfo{author}{H.~Adam}, \bibinfo{title}{Encoder-decoder with atrous
  separable convolution for semantic image segmentation}, in:
  \bibinfo{booktitle}{Proceedings of the European conference on computer vision
  (ECCV)}, \bibinfo{pages}{801--818}, \bibinfo{year}{2018}{\natexlab{c}}.

\bibitem[{Fourure et~al.(2017)Fourure, Emonet, Fromont, Muselet, Tremeau, and
  Wolf}]{fourure2017residual}
\bibinfo{author}{D.~Fourure}, \bibinfo{author}{R.~Emonet},
  \bibinfo{author}{E.~Fromont}, \bibinfo{author}{D.~Muselet},
  \bibinfo{author}{A.~Tremeau}, \bibinfo{author}{C.~Wolf},
  \bibinfo{title}{Residual Conv-Deconv Grid Network for Semantic Segmentation},
  in: \bibinfo{booktitle}{BMVC 2017}, \bibinfo{year}{2017}.

\bibitem[{Saxena and Verbeek(2016)}]{saxena2016convolutional}
\bibinfo{author}{S.~Saxena}, \bibinfo{author}{J.~Verbeek},
  \bibinfo{title}{Convolutional Neural Fabrics}, in:
  \bibinfo{booktitle}{Advances in Neural Information Processing Systems},
  \bibinfo{pages}{1--9}, \bibinfo{year}{2016}.

\bibitem[{Xie et~al.(2018)Xie, Wang, Zhang, Lai, Hong, and
  Qi}]{xie2018interleaved}
\bibinfo{author}{G.~Xie}, \bibinfo{author}{J.~Wang},
  \bibinfo{author}{T.~Zhang}, \bibinfo{author}{J.~Lai},
  \bibinfo{author}{R.~Hong}, \bibinfo{author}{G.-J. Qi},
  \bibinfo{title}{Interleaved structured sparse convolutional neural networks},
  in: \bibinfo{booktitle}{Proceedings of the IEEE Conference on Computer Vision
  and Pattern Recognition}, \bibinfo{pages}{8847--8856}, \bibinfo{year}{2018}.

\bibitem[{Zhou et~al.(2019)Zhou, Siddiquee, Tajbakhsh, and
  Liang}]{zhou2019unetplusplus}
\bibinfo{author}{Z.~Zhou}, \bibinfo{author}{M.~M.~R. Siddiquee},
  \bibinfo{author}{N.~Tajbakhsh}, \bibinfo{author}{J.~Liang},
  \bibinfo{title}{UNet++: Redesigning Skip Connections to Exploit Multiscale
  Features in Image Segmentation}, \bibinfo{journal}{IEEE Transactions on
  Medical Imaging} .

\bibitem[{Zhou et~al.(2018{\natexlab{a}})Zhou, Siddiquee, Tajbakhsh, and
  Liang}]{zhou2018unetplusplus}
\bibinfo{author}{Z.~Zhou}, \bibinfo{author}{M.~M.~R. Siddiquee},
  \bibinfo{author}{N.~Tajbakhsh}, \bibinfo{author}{J.~Liang},
  \bibinfo{title}{Unet++: A Nested U-Net Architecture for Medical Image
  Segmentation}, in: \bibinfo{booktitle}{Deep Learning in Medical Image
  Analysis and Multimodal Learning for Clinical Decision Support},
  \bibinfo{publisher}{Springer}, \bibinfo{pages}{3--11},
  \bibinfo{year}{2018}{\natexlab{a}}.

\bibitem[{Wang et~al.(2020)Wang, Sun, Cheng, Jiang, Deng, Zhao, Liu, Mu, Tan,
  Wang et~al.}]{wang2020deep}
\bibinfo{author}{J.~Wang}, \bibinfo{author}{K.~Sun},
  \bibinfo{author}{T.~Cheng}, \bibinfo{author}{B.~Jiang},
  \bibinfo{author}{C.~Deng}, \bibinfo{author}{Y.~Zhao},
  \bibinfo{author}{D.~Liu}, \bibinfo{author}{Y.~Mu}, \bibinfo{author}{M.~Tan},
  \bibinfo{author}{X.~Wang}, et~al., \bibinfo{title}{Deep high-resolution
  representation learning for visual recognition}, \bibinfo{journal}{IEEE
  transactions on pattern analysis and machine intelligence} .

\bibitem[{Hubara et~al.(2016)Hubara, Courbariaux, Soudry, El-Yaniv, and
  Bengio}]{hubara2016binarized}
\bibinfo{author}{I.~Hubara}, \bibinfo{author}{M.~Courbariaux},
  \bibinfo{author}{D.~Soudry}, \bibinfo{author}{R.~El-Yaniv},
  \bibinfo{author}{Y.~Bengio}, \bibinfo{title}{Binarized neural networks}, in:
  \bibinfo{booktitle}{Advances in Neural Information Processing Systems},
  \bibinfo{pages}{4107--4115}, \bibinfo{year}{2016}.

\bibitem[{Hubara et~al.(2017)Hubara, Courbariaux, Soudry, El-Yaniv, and
  Bengio}]{hubara2017quantized}
\bibinfo{author}{I.~Hubara}, \bibinfo{author}{M.~Courbariaux},
  \bibinfo{author}{D.~Soudry}, \bibinfo{author}{R.~El-Yaniv},
  \bibinfo{author}{Y.~Bengio}, \bibinfo{title}{Quantized neural networks:
  Training neural networks with low precision weights and activations},
  \bibinfo{journal}{Journal of Machine Learning Research}
  \bibinfo{volume}{18}~(\bibinfo{number}{1}) (\bibinfo{year}{2017})
  \bibinfo{pages}{6869--6898}.

\bibitem[{Micikevicius et~al.(2018)Micikevicius, Narang, Alben, Diamos, Elsen,
  Garcia, Ginsburg, Houston, Kuchaiev, Venkatesh, and
  Wu}]{micikevicius2018mixed}
\bibinfo{author}{P.~Micikevicius}, \bibinfo{author}{S.~Narang},
  \bibinfo{author}{J.~Alben}, \bibinfo{author}{G.~Diamos},
  \bibinfo{author}{E.~Elsen}, \bibinfo{author}{D.~Garcia},
  \bibinfo{author}{B.~Ginsburg}, \bibinfo{author}{M.~Houston},
  \bibinfo{author}{O.~Kuchaiev}, \bibinfo{author}{G.~Venkatesh},
  \bibinfo{author}{H.~Wu}, \bibinfo{title}{Mixed Precision Training}, in:
  \bibinfo{booktitle}{International Conference on Learning Representations},
  \bibinfo{year}{2018}.

\bibitem[{Kalamkar et~al.(2019)Kalamkar, Mudigere, Mellempudi, Das, Banerjee,
  Avancha, Vooturi, Jammalamadaka, Huang, Yuen et~al.}]{kalamkar2019study}
\bibinfo{author}{D.~Kalamkar}, \bibinfo{author}{D.~Mudigere},
  \bibinfo{author}{N.~Mellempudi}, \bibinfo{author}{D.~Das},
  \bibinfo{author}{K.~Banerjee}, \bibinfo{author}{S.~Avancha},
  \bibinfo{author}{D.~T. Vooturi}, \bibinfo{author}{N.~Jammalamadaka},
  \bibinfo{author}{J.~Huang}, \bibinfo{author}{H.~Yuen}, et~al.,
  \bibinfo{title}{A Study of BFLOAT16 for Deep Learning Training},
  \bibinfo{journal}{arXiv preprint arXiv:1905.12322} .

\bibitem[{Yang et~al.(2019)Yang, Zhang, Kirichenko, Bai, Wilson, and
  De~Sa}]{yang2019swalp}
\bibinfo{author}{G.~Yang}, \bibinfo{author}{T.~Zhang},
  \bibinfo{author}{P.~Kirichenko}, \bibinfo{author}{J.~Bai},
  \bibinfo{author}{A.~G. Wilson}, \bibinfo{author}{C.~De~Sa},
  \bibinfo{title}{SWALP: Stochastic Weight Averaging in Low Precision
  Training}, in: \bibinfo{booktitle}{International Conference on Machine
  Learning}, \bibinfo{pages}{7015--7024}, \bibinfo{year}{2019}.

\bibitem[{Dong et~al.(2019)Dong, Yao, Gholami, Mahoney, and
  Keutzer}]{dong2019hawq}
\bibinfo{author}{Z.~Dong}, \bibinfo{author}{Z.~Yao},
  \bibinfo{author}{A.~Gholami}, \bibinfo{author}{M.~W. Mahoney},
  \bibinfo{author}{K.~Keutzer}, \bibinfo{title}{Hawq: Hessian aware
  quantization of neural networks with mixed-precision}, in:
  \bibinfo{booktitle}{Proceedings of the IEEE International Conference on
  Computer Vision}, \bibinfo{pages}{293--302}, \bibinfo{year}{2019}.

\bibitem[{Cheng et~al.(2019)Cheng, Wang, Pan, and
  Lukasiewicz}]{cheng2019distributed}
\bibinfo{author}{Z.~Cheng}, \bibinfo{author}{W.~Wang},
  \bibinfo{author}{Y.~Pan}, \bibinfo{author}{T.~Lukasiewicz},
  \bibinfo{title}{Distributed Low Precision Training Without Mixed Precision},
  \bibinfo{journal}{arXiv preprint arXiv:1911.07384} .

\bibitem[{Simard et~al.(2003)Simard, Steinkraus, Platt et~al.}]{simard2003best}
\bibinfo{author}{P.~Y. Simard}, \bibinfo{author}{D.~Steinkraus},
  \bibinfo{author}{J.~C. Platt}, et~al., \bibinfo{title}{Best practices for
  convolutional neural networks applied to visual document analysis}, in:
  \bibinfo{booktitle}{Proceedings of the International Conference on Document
  Analysis and Recognition}, vol.~\bibinfo{volume}{2},
  \bibinfo{pages}{958--962}, \bibinfo{year}{2003}.

\bibitem[{Deng et~al.(2009)Deng, Dong, Socher, Li, Li, and
  Fei-Fei}]{deng2009imagenet}
\bibinfo{author}{J.~Deng}, \bibinfo{author}{W.~Dong},
  \bibinfo{author}{R.~Socher}, \bibinfo{author}{L.-J. Li},
  \bibinfo{author}{K.~Li}, \bibinfo{author}{L.~Fei-Fei},
  \bibinfo{title}{Imagenet: A large-scale hierarchical image database}, in:
  \bibinfo{booktitle}{Proceedings of the IEEE Conference on Computer Vision and
  Pattern Recognition}, \bibinfo{pages}{248--255}, \bibinfo{year}{2009}.

\bibitem[{Perez and Wang(2017)}]{perez2017effectiveness}
\bibinfo{author}{L.~Perez}, \bibinfo{author}{J.~Wang}, \bibinfo{title}{The
  effectiveness of data augmentation in image classification using deep
  learning}, \bibinfo{journal}{arXiv preprint arXiv:1712.04621} .

\bibitem[{Hauberg et~al.(2016)Hauberg, Freifeld, Larsen, Fisher, and
  Hansen}]{hauberg2016dreaming}
\bibinfo{author}{S.~Hauberg}, \bibinfo{author}{O.~Freifeld},
  \bibinfo{author}{A.~B.~L. Larsen}, \bibinfo{author}{J.~Fisher},
  \bibinfo{author}{L.~Hansen}, \bibinfo{title}{Dreaming more data:
  Class-dependent distributions over diffeomorphisms for learned data
  augmentation}, in: \bibinfo{booktitle}{Artificial Intelligence and
  Statistics}, \bibinfo{pages}{342--350}, \bibinfo{year}{2016}.

\bibitem[{Salehinejad et~al.(2018)Salehinejad, Valaee, Dowdell, and
  Barfett}]{salehinejad2018image}
\bibinfo{author}{H.~Salehinejad}, \bibinfo{author}{S.~Valaee},
  \bibinfo{author}{T.~Dowdell}, \bibinfo{author}{J.~Barfett},
  \bibinfo{title}{Image augmentation using radial transform for training deep
  neural networks}, in: \bibinfo{booktitle}{Proceedings of the IEEE
  International Conference on Acoustics, Speech and Signal Processing},
  \bibinfo{pages}{3016--3020}, \bibinfo{year}{2018}.

\bibitem[{Zhao et~al.(2019)Zhao, Balakrishnan, Durand, Guttag, and
  Dalca}]{zhao2019data}
\bibinfo{author}{A.~Zhao}, \bibinfo{author}{G.~Balakrishnan},
  \bibinfo{author}{F.~Durand}, \bibinfo{author}{J.~V. Guttag},
  \bibinfo{author}{A.~V. Dalca}, \bibinfo{title}{Data augmentation using
  learned transformations for one-shot medical image segmentation}, in:
  \bibinfo{booktitle}{Proceedings of the IEEE Conference on Computer Vision and
  Pattern Recognition}, \bibinfo{pages}{8543--8553}, \bibinfo{year}{2019}.

\bibitem[{Pandey et~al.(2020)Pandey, Singh, and Tian}]{pandey2020image}
\bibinfo{author}{S.~Pandey}, \bibinfo{author}{P.~R. Singh},
  \bibinfo{author}{J.~Tian}, \bibinfo{title}{An image augmentation approach
  using two-stage generative adversarial network for nuclei image
  segmentation}, \bibinfo{journal}{Biomedical Signal Processing and Control}
  \bibinfo{volume}{57} (\bibinfo{year}{2020}) \bibinfo{pages}{101782}.

\bibitem[{Dolz et~al.(2018{\natexlab{b}})Dolz, Desrosiers, and
  Ayed}]{dolz20183d}
\bibinfo{author}{J.~Dolz}, \bibinfo{author}{C.~Desrosiers},
  \bibinfo{author}{I.~B. Ayed}, \bibinfo{title}{3D fully convolutional networks
  for subcortical segmentation in MRI: A large-scale study},
  \bibinfo{journal}{NeuroImage} \bibinfo{volume}{170}
  (\bibinfo{year}{2018}{\natexlab{b}}) \bibinfo{pages}{456--470}.

\bibitem[{Sun et~al.(2019{\natexlab{b}})Sun, Xiao, Liu, and Wang}]{sun2019deep}
\bibinfo{author}{K.~Sun}, \bibinfo{author}{B.~Xiao}, \bibinfo{author}{D.~Liu},
  \bibinfo{author}{J.~Wang}, \bibinfo{title}{Deep High-Resolution
  Representation Learning for Human Pose Estimation}, in:
  \bibinfo{booktitle}{Proceedings of the IEEE Conference on Computer Vision and
  Pattern Recognition}, \bibinfo{pages}{5693--5703},
  \bibinfo{year}{2019}{\natexlab{b}}.

\bibitem[{Sun et~al.(2019{\natexlab{c}})Sun, Zhao, Jiang, Cheng, Xiao, Liu, Mu,
  Wang, Liu, and Wang}]{sun2019high}
\bibinfo{author}{K.~Sun}, \bibinfo{author}{Y.~Zhao},
  \bibinfo{author}{B.~Jiang}, \bibinfo{author}{T.~Cheng},
  \bibinfo{author}{B.~Xiao}, \bibinfo{author}{D.~Liu}, \bibinfo{author}{Y.~Mu},
  \bibinfo{author}{X.~Wang}, \bibinfo{author}{W.~Liu},
  \bibinfo{author}{J.~Wang}, \bibinfo{title}{High-Resolution Representations
  for Labeling Pixels and Regions}, \bibinfo{journal}{arXiv preprint
  arXiv:1904.04514} .

\bibitem[{Ulyanov et~al.(2016)Ulyanov, Vedaldi, and
  Lempitsky}]{ulyanov2016instance}
\bibinfo{author}{D.~Ulyanov}, \bibinfo{author}{A.~Vedaldi},
  \bibinfo{author}{V.~Lempitsky}, \bibinfo{title}{Instance normalization: The
  missing ingredient for fast stylization}, \bibinfo{journal}{arXiv preprint
  arXiv:1607.08022} .

\bibitem[{Cornea(2009)}]{cornea2009ieee}
\bibinfo{author}{M.~Cornea}, \bibinfo{title}{IEEE 754-2008 Decimal
  Floating-Point for Intel{\textregistered} Architecture Processors}, in:
  \bibinfo{booktitle}{IEEE Symposium on Computer Arithmetic},
  \bibinfo{pages}{225--228}, \bibinfo{year}{2009}.

\bibitem[{Shattuck et~al.(2008)Shattuck, Mirza, Adisetiyo, Hojatkashani,
  Salamon, Narr, Poldrack, Bilder, and Toga}]{shattuck2008construction}
\bibinfo{author}{D.~W. Shattuck}, \bibinfo{author}{M.~Mirza},
  \bibinfo{author}{V.~Adisetiyo}, \bibinfo{author}{C.~Hojatkashani},
  \bibinfo{author}{G.~Salamon}, \bibinfo{author}{K.~L. Narr},
  \bibinfo{author}{R.~A. Poldrack}, \bibinfo{author}{R.~M. Bilder},
  \bibinfo{author}{A.~W. Toga}, \bibinfo{title}{Construction of a 3D
  probabilistic atlas of human cortical structures},
  \bibinfo{journal}{NeuroImage} \bibinfo{volume}{39}~(\bibinfo{number}{3})
  (\bibinfo{year}{2008}) \bibinfo{pages}{1064--1080}.

\bibitem[{Tustison et~al.(2010)Tustison, Avants, Cook, Zheng, Egan, Yushkevich,
  and Gee}]{tustison2010n4itk}
\bibinfo{author}{N.~J. Tustison}, \bibinfo{author}{B.~B. Avants},
  \bibinfo{author}{P.~A. Cook}, \bibinfo{author}{Y.~Zheng},
  \bibinfo{author}{A.~Egan}, \bibinfo{author}{P.~A. Yushkevich},
  \bibinfo{author}{J.~C. Gee}, \bibinfo{title}{N4ITK: improved N3 bias
  correction}, \bibinfo{journal}{IEEE Transactions on Medical Imaging}
  \bibinfo{volume}{29}~(\bibinfo{number}{6}) (\bibinfo{year}{2010})
  \bibinfo{pages}{1310--1320}.

\bibitem[{Avants et~al.(2009)Avants, Tustison, and Song}]{avants2009advanced}
\bibinfo{author}{B.~B. Avants}, \bibinfo{author}{N.~Tustison},
  \bibinfo{author}{G.~Song}, \bibinfo{title}{Advanced normalization tools
  (ANTS)}, \bibinfo{journal}{Insight j}
  \bibinfo{volume}{2}~(\bibinfo{number}{365}) (\bibinfo{year}{2009})
  \bibinfo{pages}{1--35}.

\bibitem[{Paszke et~al.(2019)Paszke, Gross, Massa, Lerer, Bradbury, Chanan,
  Killeen, Lin, Gimelshein, Antiga et~al.}]{paszke2019pytorch}
\bibinfo{author}{A.~Paszke}, \bibinfo{author}{S.~Gross},
  \bibinfo{author}{F.~Massa}, \bibinfo{author}{A.~Lerer},
  \bibinfo{author}{J.~Bradbury}, \bibinfo{author}{G.~Chanan},
  \bibinfo{author}{T.~Killeen}, \bibinfo{author}{Z.~Lin},
  \bibinfo{author}{N.~Gimelshein}, \bibinfo{author}{L.~Antiga}, et~al.,
  \bibinfo{title}{Pytorch: An imperative style, high-performance deep learning
  library}, in: \bibinfo{booktitle}{Advances in Neural Information Processing
  Systems}, \bibinfo{pages}{8026--8037}, \bibinfo{year}{2019}.

\bibitem[{Liu et~al.(2020)Liu, Jiang, He, Chen, Liu, Gao, and Han}]{Liu2020On}
\bibinfo{author}{L.~Liu}, \bibinfo{author}{H.~Jiang}, \bibinfo{author}{P.~He},
  \bibinfo{author}{W.~Chen}, \bibinfo{author}{X.~Liu},
  \bibinfo{author}{J.~Gao}, \bibinfo{author}{J.~Han}, \bibinfo{title}{On the
  Variance of the Adaptive Learning Rate and Beyond}, in:
  \bibinfo{booktitle}{International Conference on Learning Representations},
  \bibinfo{year}{2020}.

\bibitem[{Wu et~al.(2015)Wu, Kim, Sanroma, Wang, Munsell, Shen, Initiative
  et~al.}]{wu2015hierarchical}
\bibinfo{author}{G.~Wu}, \bibinfo{author}{M.~Kim},
  \bibinfo{author}{G.~Sanroma}, \bibinfo{author}{Q.~Wang},
  \bibinfo{author}{B.~C. Munsell}, \bibinfo{author}{D.~Shen},
  \bibinfo{author}{A.~D.~N. Initiative}, et~al., \bibinfo{title}{Hierarchical
  multi-atlas label fusion with multi-scale feature representation and
  label-specific patch partition}, \bibinfo{journal}{NeuroImage}
  \bibinfo{volume}{106} (\bibinfo{year}{2015}) \bibinfo{pages}{34--46}.

\bibitem[{Kervadec et~al.(2019)Kervadec, Bouchtiba, Desrosiers, Granger, Dolz,
  and Ayed}]{kervadec2019boundary}
\bibinfo{author}{H.~Kervadec}, \bibinfo{author}{J.~Bouchtiba},
  \bibinfo{author}{C.~Desrosiers}, \bibinfo{author}{E.~Granger},
  \bibinfo{author}{J.~Dolz}, \bibinfo{author}{I.~B. Ayed},
  \bibinfo{title}{Boundary loss for highly unbalanced segmentation}, in:
  \bibinfo{booktitle}{International Conference on Medical Imaging with Deep
  Learning}, \bibinfo{pages}{285--296}, \bibinfo{year}{2019}.

\bibitem[{Zhou et~al.(2018{\natexlab{b}})Zhou, Xu, and Gu}]{zhou2018stochastic}
\bibinfo{author}{D.~Zhou}, \bibinfo{author}{P.~Xu}, \bibinfo{author}{Q.~Gu},
  \bibinfo{title}{Stochastic nested variance reduction for nonconvex
  optimization}, in: \bibinfo{booktitle}{Advances in Neural Information
  Processing Systems}, \bibinfo{pages}{3921--3932},
  \bibinfo{year}{2018}{\natexlab{b}}.

\bibitem[{Nakkiran et~al.(2020)Nakkiran, Kaplun, Bansal, Yang, Barak, and
  Sutskever}]{Nakkiran2020Deep}
\bibinfo{author}{P.~Nakkiran}, \bibinfo{author}{G.~Kaplun},
  \bibinfo{author}{Y.~Bansal}, \bibinfo{author}{T.~Yang},
  \bibinfo{author}{B.~Barak}, \bibinfo{author}{I.~Sutskever},
  \bibinfo{title}{Deep Double Descent: Where Bigger Models and More Data Hurt},
  in: \bibinfo{booktitle}{International Conference on Learning
  Representations}, \bibinfo{year}{2020}.

\end{thebibliography}
\bibliographystyle{elsarticle-num-names}

\end{document}